\documentclass{aa}
\usepackage[varg]{txfonts}
\usepackage{graphicx}
\graphicspath{{figures/}}
\usepackage{natbib}
\bibpunct{(}{)}{;}{a}{}{,} 
\usepackage{hyperref}
\hypersetup{
	colorlinks=true,
	linkcolor=blue,
	citecolor=blue
}

\begin{document}

\title{The dependence of intrinsic alignment of galaxies on wavelength using KiDS and GAMA}

\author{Christos Georgiou \inst{1}\thanks{georgiou@strw.leidenuniv.nl}
	\and Harry Johnston \inst{2}
	\and Henk Hoekstra \inst{1}
	\and Massimo Viola \inst{1}
	\and Konrad Kuijken \inst{1}
	\and Benjamin Joachimi \inst{2}
	\and Nora Elisa Chisari \inst{3}
	\and Daniel J. Farrow \inst{4}
	\and Hendrik Hildebrandt \inst{5}
	\and Benne W. Holwerda \inst{6}
	\and Arun Kannawadi \inst{1}}
\institute{Leiden Observatory, Leiden University, Niels Bohrweg 2, 2333 CA Leiden, The Netherlands
	\and Department of Physics and Astronomy, University College London, Gower Street, WC1E 6BT London, UK
	\and Department of Physics, University of Oxford, Keble Road, Oxford OX1 3RH, UK
	\and Max-Planck-Institut fuer extraterrestrische Physik, Postfach 1312 Giessenbachstrasse, D-85741 Garching, Germany
	\and Argelander-Institut f\"ur Astronomie, Auf dem H\"ugel 71, 53121 Bonn, Germany
	\and Department of Physics and Astronomy, 102 Natural Science Building, University of Louisville, Louisville KY 40292, USA}

\date{Received <date> / Accepted <date>}

\abstract{The outer regions of galaxies are more susceptible to the tidal interactions that lead to intrinsic alignments of galaxies. The resulting alignment signal may therefore depend on the passband if the colours of galaxies vary spatially. To quantify this, we measured the shapes of galaxies with spectroscopic redshifts from the GAMA survey using deep $gri$ imaging data from the KiloDegree Survey. The performance of the moment-based shape measurement algorithm DEIMOS was assessed using dedicated image simulations, which showed that the ellipticities could be determined with an accuracy better than 1\% in all bands. Additional tests for potential systematic errors did not reveal any issues. We measure a significant difference of the alignment signal between the $g,r$ and $i$-band observations. This difference exceeds the amplitude of the linear alignment model on scales below 2 Mpc$/h$. Separating the sample into central/satellite and red/blue galaxies, we find that the difference is dominated by red satellite galaxies.}


\maketitle

\section{Introduction}
\label{sec:Introduction}

In the last few decades advances in studying the cosmos have led to a regime of ``precision cosmology''. 
New probes and techniques have significantly increased the sensitivity with which we can measure cosmological parameters that describe our Universe \citep[see e.g.][]{Planck2015,BOSS,PanSTARRS1}. This has set the foundations for the establishment of the $\Lambda$CDM model as the concordance cosmological model. However, this standard model of cosmology is surrounded by a big mystery: we are not at all certain as to what dark matter and dark energy consist of, components that make up approximately 95\% of the Universe's energy density at the present epoch. The standard $\Lambda$CDM model assumes general relativity with $\Lambda$, a cosmological constant, to explain the Universe's accelerated expansion, but this $\Lambda$ is extremely small compared to the vacuum energy expected from quantum field theory. In addition, the interpretation of this cosmological constant is a missing piece of the standard model. To address this problem one can introduce a new dark energy fluid \citep[see e.g.][]{DarkEnergyReview}, whose nature remains an open question, or modify the equations of General Relativity to explain the late time accelerated expansion of the Universe \citep[for a review see][]{Koyama}.

Studying the ``dark sector'' of the Universe is critical in solving this mystery, but is also very challenging. Dark energy models and modified gravity theories are very hard to distinguish among themselves, and dark matter is ``invisible'' since it interacts only gravitationally with baryonic matter. The first problem can be tackled with the acquisition of more, higher quality data. As for the second problem, weak gravitational lensing has proven to be a powerful tool \citep[for a review see][]{Bartlemann2001}. Coherent distortion of light rays is caused by the matter between the source and the observer, and it can be used to study dark matter directly \citep[e.g.][]{BulletCluster, DarkMatterLensing}. In addition, weak gravitational lensing is also sensitive to the geometry of the Universe making it a powerful tool for constraining cosmology and gravity \citep[e.g.][]{Hoekstra2008, Kilbinger2015}.

Weak lensing changes the observed ellipticity of galaxies at the per cent level which is much smaller than variations in intrinsic ellipticities between galaxies, and as a result, the lensing distortion patterns can only be observed statistically, by correlating shapes of a large ensemble of galaxies. Many ongoing surveys such as the Kilo Degree Survey\footnote{http://kids.strw.leidenuniv.nl/}, the Dark Energy Survey\footnote{https://www.darkenergysurvey.org/} and the Hyper Suprime Cam survey\footnote{http://hsc.mtk.nao.ac.jp/ssp/}, as well as upcoming surveys such as \emph{Euclid}\footnote{https://www.euclid-ec.org/} and the Large Synoptic Survey Telescope\footnote{https://www.lsst.org/}, aim to exploit the phenomenon and measure cosmological parameters to very high precision \citep[e.g.][]{Hildebrandt2016,DESY1cosmicshear}. The statistical power of future surveys is high enough that the systematic uncertainty in measured shapes needs to be ultimately controlled to permille precision (e.g. \citealt{Massey2013} for \emph{Euclid}). 

If galaxies are intrinsically randomly oriented in the sky, any shape correlation observed could be attributed solely to gravitational lensing. However, shape correlations can also be induced during structure formation, since large-scale tidal gravitational fields affect the orientation of galaxies with respect to the matter density field, a phenomenon called intrinsic alignment. Physically associated galaxies form and evolve in similar gravitational fields, hence they are coherently aligned to some extent \citep[e.g.][]{Joachimi2015,Troxel2015}. Consequently, intrinsic alignments are a major astrophysical contaminant of weak lensing and modelling the effect is of crucial importance for high precision weak lensing measurements.

Intrinsic alignments have been studied through cosmological numerical simulations. These have revealed that dark matter haloes tend to align with each other and the matter density field and that red galaxies tend to align with red centrals \citep[for a review see][]{Kiessling2015}. This picture has been broadly confirmed by observations  \citep[e.g.][]{Mandelbaum2006,Hirata2007,Okumura2009,Joachimi2011,Li2013,Singh2015a}. Interestingly, alignments between blue galaxies have not yet been firmly detected \citep{WiggleZalignments,Heymans2013} while luminous red galaxies give a significant alignment signal.

Another important characteristic of the intrinsic alignment signal is its dependence on the galaxy's radial scale. By their nature, tidal interactions have a stronger impact on the outer parts of a galaxy than the inner ones \citep{Kormendy}. Since intrinsic alignments are attributed to tidal gravitational fields, one can expect that measuring shapes of galaxies at larger radii would give a stronger alignment signal. This dependence has been established using several cosmological hydrodynamical simulations where the alignment signal was measured using two different shape estimators: the signal was weaker for estimators that up-weighted the inner regions of a galaxy and down-weight the outer ones \citep{Chisari2015,Marco,Hilbert}.

Evidence for such a dependence in the alignment of galaxies is also provided by observations. \citet{Singh} used a galaxy sample with shapes from three different estimators to measure intrinsic alignments. The amplitude of the signal was lower for shape measurements that give more weight to the inner parts of galaxies. Similar results were seen for the alignment of central galaxies with the group satellites \citep{Huang1} as well as for the radial alignment of satellites with respect to their BCG \citep{Huang2}. However, these results were based on shapes from different shape estimators, which are sensitive to systematic uncertainties in different ways, so drawing a firm conclusion is difficult.  Ideally, one would want to measure the alignment signal using a consistent shape measurement method that can be adjusted to measure shapes from different galaxy scales, without introducing further systematic errors.

A complication in the interpretation is that in general, both spiral and elliptical galaxies appear to have outer regions that are bluer than the inner ones \citep{deJong, Franx, Peletier}, a result mainly attributed to radial gradients in the stellar populations \citep[e.g.][]{Tortora}. Since the outer regions of galaxies are more luminous in blue filters, we expect the sizes of these galaxies to be larger in these filters as well \citep{MacArthur2003}. Thus, observing in blue filters gives more weight to the outer regions of galaxies, while red filter observations are mostly revealing their inner regions. This can potentially lead to a difference in the intrinsic alignment signal as measured from different broad band images, with blue filters exhibiting a stronger signal than red ones, for a given galaxy. 

In this work we measure differences in the intrinsic alignment signal obtained from different broad band filter observations by combining data from the Galaxy And Mass Assembly survey\footnote{http://www.gama-survey.org} \citep[GAMA;][]{Driver2009,Driver2011,Liske2015} and the Kilo Degree Survey \citep[KiDS;][]{deJong2015,deJong2017}. The latter is a deep imaging survey designed primarily for weak gravitational lensing science, providing imaging data of exquisite quality. We measure galaxy ellipticities in $gri$ broad band imaging data using the same shape measurement estimator, \textsc{DEIMOS} \citep{DEIMOS}, a moment-based method that is described in Sect. \ref{sec:DEIMOS}. This shape measurement method includes an exact treatment of the point-spread function (PSF) and a correction for the weighting scheme introduced in measuring galaxy brightness. Moreover, this work is an extension to the shape catalogues already produced by the KiDS team, which do not include galaxies with magnitude $r<20$. The data used are described in Sect. \ref{sec:Data}, and in Sect. \ref{sec:PSF model} we present the PSF modelling. To calibrate our shape measurements, we use dedicated image simulations, outlined in Sect. \ref{sec:simulations}, trying to accurately replicate our galaxy sample's properties. Our results are presented in Sect. \ref{sec:results}, followed by summary and discussion in Sect. \ref{sec:Discussion}. Throughout this work, we assume a flat $\Lambda$CDM cosmology with $h=0.7$ and $\Omega_m=0.25$, to be consistent with \citet{Johnston}, as well as previous works on galaxy intrinsic alignments.

\section{The DEIMOS shape measurement method}
\label{sec:DEIMOS}

Measuring accurate shapes of galaxies from optical astronomical images is a non-trivial task. Among other things, one needs to account for the presence of noise in the data and the distortion caused by the point-spread function (PSF), and these are treated differently by different shape measuring methods. The method that we chose to use is \textsc{DEIMOS} \citep{DEIMOS}, which stands for DEconvolution In MOment Space. This is a moment-based method, meaning that surface brightness moments are calculated from image data to extract shape information of galaxies. It is an improvement over some similar approaches, such as the KSB method \citep{KSB}, because the moments of the galaxies are corrected exactly for the effect of the PSF. In addition, the effect of the weighting function employed when measuring galaxy moments (which is required to suppress the noise) is compensated using measurements of higher order moments of the galaxies. Since DEIMOS is moment-based, no assumption is made for a galaxy's model and morphology, it is much faster than model-fitting shape measurement methods and also allows flexibility in varying the weighting function, consequently enabling us to measure shapes at different galactic scales. In future work, we aim to use this flexibility to probe directly the dependence of the intrinsic alignment signal on the galaxy scales probed. \citet{DEIMOS} demonstrated the high accuracy of the method, using image simulations of the GREAT08 challenge \citep{GREAT08}. 

The (unweighted) moments of the surface brightness distribution $G(\mathbf{x})$ of a galaxy are expressed by the integral
\begin{equation}
Q_{ij}\equiv \{G\}_{i,j}=\int G(\mathbf{x})\, x_1^i\, x_2^j\, \rm{d}\mathbf{x}\,,
\label{eq:unweighted moments}
\end{equation}
where $\mathbf{x}=\{x_1, x_2\}$ are the Cartesian coordinates with the galaxy's centroid at the origin. The second order moments can be combined to estimate the ellipticity of the object:
\begin{equation}
\epsilon\equiv \epsilon_1+{\rm i}\epsilon_2=\frac{Q_{20}-Q_{02}+{\rm i}2Q_{11}}{Q_{20}+Q_{02}+2\sqrt{Q_{20}Q_{02}-Q_{11}^2}}\,,
\label{eq:ellipticity}
\end{equation}
where $|\epsilon|$ is related to the semi-minor to semi-major axis ratio $q$ of by $|\epsilon|=(1-q)/(1+q)$. 
\subsection{Distortion due to the PSF}

Images obtained with an optical telescope undergo a series of transformations that alter the observed brightness profile of objects and complicate the measurement of the galaxy shapes. One inevitable example is the effect of the PSF on the image, which is caused by the atmospheric blurring (for ground-based observations) and the optics of the telescope. The object's surface brightness $G(\mathbf{x})$ is not directly accessible in image data because it is convolved with the PSF kernel $P(\mathbf{x})$ and in practice what we observe is the PSF-convolved image,
\begin{equation}
G^*(\mathbf{x})=\int G(\mathbf{x}') \,P(\mathbf{x-x}')\,\rm{d}\mathbf{x}'
\label{eq:PSF convolution}\,.
\end{equation}

This convolution smears and distorts the true object and needs to be accounted for in order to accurately retrieve shape information. \citet{DEIMOS} showed that the moments of the \emph{true} surface brightness of the galaxy $G$ are related to those of the observed $G^*$ and the PSF kernel $P$, through
\begin{equation}
\{G^*\}_{i,j}=\sum_k^i \sum_l ^j \binom{i}{k} \binom{j}{l} \{G\}_{k,l}\{P\}_{i-k,j-k}\,.
\label{eq:Deconvolution}
\end{equation}
In order to calculate the unweighted moments of $G$ up to 2nd order all we need to know are the moments of the image $G^*$ as well as the moments of $P$ up to the same order. This does not impose any prior assumption on the profile of the PSF, contrary to some other moment-based methods such as the KSB \citep{KSB} and re-Gaussianization \citep{reGauss} methods. The moments of the PSF can be calculated by modelling the PSF, and we describe our modelling in Sect. \ref{sec:PSF model}.

\subsection{Effect of noise and weighting}

Another very important feature of real images is the presence of noise, which is mainly caused by the shot noise of counting photons, the read-out process of the detector and the sky background noise. When sky-background limited, noise in astronomical images is typically Gaussian and uncorrelated between pixels. With $N(\mathbf{x})$ expressing the noise, the flux we measure in imaging data is 
\begin{equation}
I(\mathbf{x})=G^*(\mathbf{x})+N(\mathbf{x})\,.
\label{eq:noise}
\end{equation}

The second order moments of the image have a quadratic radial weighting that enhances the sensitivity of regions far from the galaxy's centre and, in the presence of noise, leads to infinite variance. In practice, we need to choose a large enough image section, or ``postage stamp''\footnote{The dimension of the postage stamp should be set by the desired accuracy. A small postage stamp can end up truncating the moments and lead to biased results (truncation bias). See Sect. \ref{sec:simulations} for more information on the choice of a postage stamp.} on which we calculate the integral of equation \eqref{eq:unweighted moments}, but the second order moments will be completely dominated by noise in regions far from the galaxy's centre. This is dealt with by applying a weight function $W(\mathbf{x})$, centred on the galaxy, in order to suppress the noise at large separations. The measured flux then becomes
\begin{equation}
I_w(\mathbf{x}) = W(\mathbf{x})\, I(\mathbf{x})\,.
\label{eq:weighting}
\end{equation}

Typically a Gaussian weight function centred on the galaxy's centroid is employed in the measurements \citep[e.g.][]{KSB}. Ideally, we want the weight function to match the shape of the galaxy, so that galaxy light is suppressed as little as possible, and its Signal-to-Noise ratio (SNR) is maximised. Since galaxy shapes are usually elliptical, it makes sense to employ an \emph{elliptical} Gaussian weight function whose centroid, size and ellipticity are matched to that of the galaxy. This is done according to the algorithm described in \citet{Bernstein+Jarvis2002}, Sect. 3.1.2, and the weight function employed in this work is
\begin{equation}
W(\mathbf{x}) = \exp\left[ - (\mathbf{x}-\mathbf{x}_c)^\mathrm{T} \begin{pmatrix}
1-\epsilon_1 & -\epsilon_2 \\
-\epsilon_2 & 1+\epsilon_1
\end{pmatrix}
\frac{(\mathbf{x}-\mathbf{x}_c)}{2r_{\rm wf}^2}
\right]\,,
\label{eq:weight_function}
\end{equation}
where $\epsilon_i$ are the ellipticity components of the weight function, $\mathbf{x}_c$ is the centroid of the galaxy and $r_{\rm wf}$ is the scale of the weight function. 

The value for $r_{\rm wf}$ is usually optimized such that the weight function matches the galaxy's surface brightness profile, thus maximizing the SNR of the measurement. However, its value also describes the physical scale for which the galaxy's shape is measured: a small $r_{\rm wf}$ will make the shape measurement sensitive to the galaxy's bulge, while a larger $r_{\rm wf}$ will make the shape measurement more sensitive to the outskirts of the galaxy.

The weight function is iteratively matched to each galaxy through the following procedure:
\begin{enumerate}
	\item First, the centroid of the galaxy is calculated by requiring the first order moments to vanish. These moments are weighted with a circular Gaussian function of size $r_{\rm wf}$.
	\item Once the centroid is determined, the second order moments (weighted with a circular Gaussian function of size $r_{\rm wf}$) are used to measure the ellipticity through Eq. \eqref{eq:ellipticity}.
	\item Then, the measured ellipticity is used to define the new, elliptical weight function with which the galaxy's shape is measured again.
	\item This procedure is repeated until the SNR\footnote{The SNR is calculated according to Eq. 3.14 of \citet{Bernstein+Jarvis2002}.} of the measurement converges. 
\end{enumerate}
The weight function's ellipticity components and centroid are determined for every galaxy individually. We also choose to preserve the area of the weight function between each matching iteration. This means that starting from a circular shape, the weight function will become more stretched along the semi-major axis of the galaxy with each iteration, and squeezed in the direction of the semi-minor axis, eventually matching the size of the galaxy. Our choice of $r_{\rm wf}$ is described in Sect. \ref{sec:simulations-weightFunction}.

This results in estimates of the weighted moments $\{I_w\}_{i,j}$; to obtain the correct shape information we need the unweighted moments of equation \eqref{eq:unweighted moments}. Employing a weight function biases the shape measurement, and to minimise this a de-weighting procedure is required. This is done by inverting equation \eqref{eq:weighting} for $I=I_w/W$ and expanding $1/W$ in a Taylor series around $\mathbf{x}_c$ (see \citealt{DEIMOS} for details). The maximum order of the Taylor expansion is a free parameter, denoted by $n_w$.
The de-weighting procedure relies on calculating higher order weighted moments and the inevitable truncation of the Taylor series expansion introduces a bias to the measured shapes, dubbed de-weighting bias. The overall bias of the shape measurement (which includes bias due to noise and de-weighting) is characterized in Sect \ref{sec:simulations-weightFunction}.

\subsection{Error and flags}
\label{sec:errorsflags}

The error calculation on the measured de-weighted moments is described in \citet{DEIMOS}. It takes advantage of the nearly linear response of the measured ellipticity dispersion to the correction order $n_w$. The covariance matrix of the weighted moments is used for the calculation of the error on the de-weighted moments. However, this matrix is sensitive to the weighting function. We observe that, for a small weight function the error on the measured ellipticity is small but the shape measurement is strongly biased as a result of using a small weight function. To avoid this discrepancy, when calculating correlation functions to measure the intrinsic alignment signal, we do not use the errors on the ellipticities. This is not expected to affect our results, because our galaxy sample is bright, with a high SNR, and the ellipticity errors are generally much smaller than the ellipticity rms.

Furthermore, there are four different flags for keeping track of problematic shape measurements. The first one is raised when the centroid determination gives a final centroid shifted by more than 5 pixels from the input one. Two flags are related to measurements of non-sensical moments or ellipticity (i.e. $Q_{00}, Q_{20}, Q_{02}<0$, $Q_{11}^2>Q_{20}Q_{02}$ or $\epsilon>1$). One of the flags is raised when the measurement is done prior to deconvolution and the other is raised for measurements after deconvolution. The fourth flag indicates that the ellipticity matching has failed. This means that during the iterative process of matching a weight function to the galaxy (described in the previous section) the SNR of the shape measurement for the different weight functions did not converge. In our analysis, we only consider shapes of galaxies that do not raise any of these flags.

\section{Data}
\label{sec:Data}

Intrinsic alignments between galaxies are believed to be caused by tidal gravitational forces and their measurement require physically associated galaxies, which can be identified as long as precise distance information is provided, usually made available with spectroscopy. Since the phenomenon is a correlation of galaxy shapes, high quality images are also needed from which shapes of galaxies can be accurately measured. The galaxy sample used in this work is obtained from the GAMA survey, which provides spectroscopic redshift information, and shapes were measured from KiDS deep imaging data. 

\subsection {GAMA}

Galaxy And Mass Assembly \citep[GAMA,][]{Driver2009, Driver2011, Liske2015} is a spectroscopic survey of $\sim300,000$ galaxies with a magnitude limit of Petrosian $r_{\rm AB}<19.8$ mag covering $\sim286$ deg$^2$ of the sky in five patches. In this work we will use the three 12$\times$5 deg$^2$ equatorial fields, centred  at approximately 9$^h$, 12$^h$ and 15$^h$ RA, named G09, G12 and G15, respectively, which contain $\sim180,000$ galaxies. One advantage of the GAMA survey is its high completeness: in the three equatorial regions the redshift completeness exceeds 98\%. This results in a clean galaxy sample and measurement, without the complications of selection effects.

\subsection{KiDS}

Kilo Degree Survey \citep[KiDS,][]{deJong2015,deJong2017} is an ongoing deep imaging survey carried out with the OmegaCAM CCD mosaic camera mounted on the VLT Survey Telescope (VST). The survey aims to cover 1350  deg$^2$ area on the sky in four SDSS-like bands ($u$, $g$, $r$ and $i$) down to a limiting magnitude of 24.3, 25.1, 24.9 and 23.8 (5$\sigma$ in a 2 arcsec aperture), respectively. As the primary science goal of the survey is cosmology with weak lensing, tight constrains are set on the observing conditions and the camera performance which guarantee a well-behaved, small and nearly round PSF.

The images used in this work are from the KiDS-450 dataset \citep{Hildebrandt2016} and they completely cover the three GAMA equatorial patches. Sub-exposures from the four bands are reduced and calibrated using the \textsc{Astro-WISE} system \citep{Astrowise1,Astrowise2}. This procedure involves de-trending of the raw images which takes care of cross-talk correction between CCDs, artefacts such as cosmic rays and flat-fielding as well as background subtraction. Next, the sub-exposures are photometrically and astrometrically calibrated and the images are co-added to produce the final image product. 

We use these co-added images in $g$, $r$ and $i$-band filters to measure the shapes of GAMA galaxies. We choose to not process the $u$-band image data because the image quality and depth are significantly lower than the other bands and the measured shapes will be harder to interpret. A main difference from the shear catalogues used in \cite{Hildebrandt2016} is the fact that our measurements are done on the \textsc{Astro-WISE} reduced images, instead of the \textsc{Theli} reduced ones (see \citealt{Kuijken2015} and references therein), because only the $r$ and $i$-band images are reduced with the \textsc{Theli} pipeline.

Since the co-added images consist of multiple dithers, the corners of adjacent image tiles will share a common section of the sky. This results in objects being imaged in 2,3 or 4 different image tiles and leaves us with multiple shape measurements of the same galaxy. To deal with this, we make use of the corresponding weight maps: cutting a postage stamp (50$\times$50 pixels) of the weight map around the galaxy's position for each image, we calculate the mean value of this postage stamp and keep the shape measurement for which the mean value of the weight map is the largest. This ensures that, for all the multiply imaged galaxies, we use the shape measurement from the ``cleanest'', highest SNR co-added image.

\section{Modelling the spatial variation of the PSF}
\label{sec:PSF model}

All astronomical observations in optical wavelengths are carried out using optical systems that alter the observed image. In addition, ground-based observations suffer from time-depended atmospheric distortions that are caused by turbulence in the atmosphere. All these effects are quantified by the PSF and shape measurement techniques need to account for this blurring in order to retrieve shapes accurately.

To correct for the distortion caused by the PSF, in the framework of \textsc{DEIMOS}, we require the calculation of up to 2nd order unweighted moments of the PSF. The PSF can be measured using stars (point sources) detected in the image. However, we are interested in the PSF at the positions of the galaxies, so interpolation of a PSF model is required. The accuracy of the interpolation depends on the star number density and distribution across the image \citep[e.g.][]{Hoekstra2004}.

A big advantage of using KiDS imaging data for shape measurements is the well behaved PSF of the images. In \cite{Kuijken2015} the smoothness of the PSF across the whole focal plane is demonstrated, which can also be seen from PSF ellipticity and size distributions in \cite{deJong2017}. The $r$-band PSF has a median full-width at half maximum (FWHM) of 0.68 arcsec while the values in the $g$ and $i$-band are 0.85 and 0.79 arcsec, respectively. The mean ellipticity is $\sim$0.05 in all three filters.

In this work, stars used for the PSF modelling are selected using \textsc{SExtractor}: unsaturated, high-SNR objects are detected and the \texttt{FLUX\_AUTO} vs. \texttt{FLUX\_RADIUS} plot is used to identify the stellar locus in an automated way, as described below:
\begin{enumerate}
	\item Firstly, the median \texttt{FLUX\_RADIUS} is calculated for the brightest objects ($f_{\rm max}/30<\texttt{FLUX\_AUTO}<f_{\rm max}$) with sizes larger than 0.5 image pixels. Here, $f_{\rm max}$ is the flux of the 20th brightest object in the image (the first 20 are excluded as saturated sources).
	\item Secondly, a new median size $r_{\rm med}$ is obtained using objects smaller than 1.5 times the previous median size and with flux larger than $f_{\rm max}/100$. 
	\item As a final step, the selected stars are objects with sizes between $r_{\rm med}/2$ and $r_{\rm med}+0.2$" and fluxes larger than $f_{\rm max}/100$. 
\end{enumerate}
This procedure manages to isolate the area of very compact and bright objects in the flux versus size plot and the resulting star density is a few thousand stars per 1x1 deg$^2$ image tile.

The selected stars are then used to model the shape of the PSF at their position using the shapelets expansion \citep{shapeletsI} and the procedure is described in Appendix A1 of \cite{Kuijken2015}\footnote{The difference being that in \cite{Kuijken2015} the different sub-exposures are used while, in this work, the modelling is done on the co-added images.}. Shapelets refer to a set of basis functions for images, constructed by the product of Gaussians and Hermite polynomials in 2-dimensions. Any 2-dimensional image can be described by a linear combination of shapelets but the accuracy depends on the maximum Hermite polynomial order. We model the shape of the PSF using Hermite polynomials of up to 10th order (set by the pixel scale of the images, following \citealt{Kuijken2015}). The spatial variation of the PSF across the image is then modelled with a 4th order polynomial and the PSF moments at the position of galaxies can be analytically computed. We chose this polynomial order to be consistent with previous analysis of KiDS data, where it has been shown that this PSF model does not produce large correlations in residuals between stars and the PSF model \cite[Fig. 5]{Kuijken2015}. We refer the read to \cite{Kuijken2015} for a mathematical description of the PSF model. 

\begin{figure}
	\resizebox{\hsize}{!}{\includegraphics{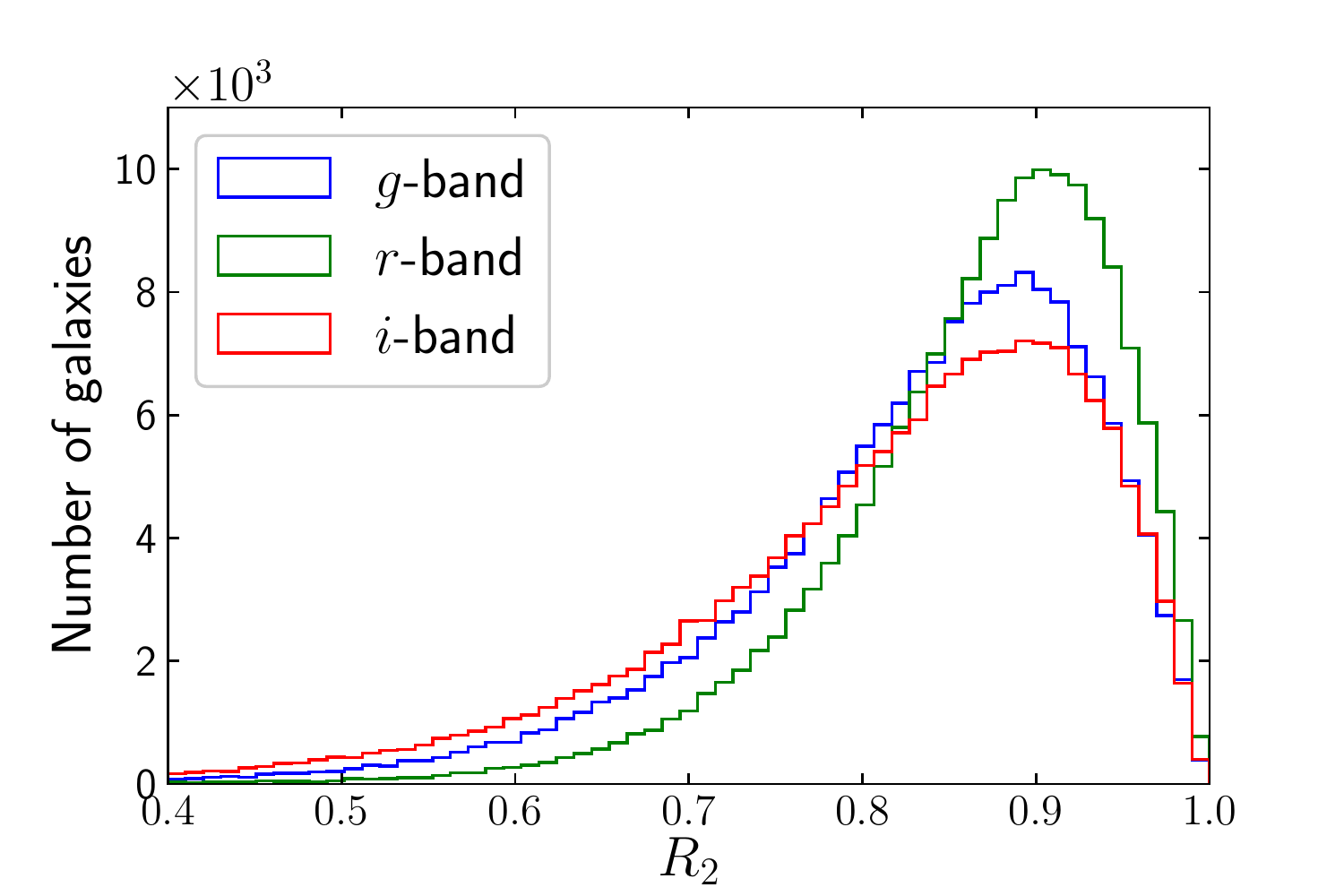}}
	\caption{Galaxy resolution distribution of GAMA galaxies for the different broad band filters as measured from KiDS image data. If the galaxy is much larger than the PSF, $R_2$ is close to 1.}
	\label{fig:R0_histograms}
\end{figure}

In Fig. \ref{fig:R0_histograms} we show the histograms of the galaxy resolution $R_2$ for our GAMA galaxy sample, as measured from KiDS imaging data. $R_2$ is defined as
\begin{equation}
R_2=1-\frac{T^{\rm PSF}}{T^{\rm gal}}\,,
\label{eq:R2}
\end{equation}
where $T^{\rm PSF}=Q_{20}^{\rm PSF}+Q_{02}^{\rm PSF}$ is the trace of the matrix of the unweighted second order moments of the PSF while $T^{\rm gal}$ makes use of the de-weighted moments of the \emph{convolved} galaxy (which are approximately its unweighted moments). $R_2$ is effectively comparing the size of the galaxy to the size of the PSF \citep{Mandelbaum2011}. A value close to 1 means the PSF is much smaller than the galaxy's size and $R_2$ close to zero means the PSF and galaxy have similar sizes. Figure \ref{fig:R0_histograms} demonstrates that GAMA galaxies are generally well resolved, being much larger than the PSF, and also shows the difference of $R_2$ in the three broad band filters: in the $r$-band galaxies are best resolved, followed by $g$ and then $i$-band.

\section{Image simulations}
\label{sec:simulations}

Measurements of galaxy ellipticities require some non-linear manipulation of the image pixel data. Because of this, PSF convolution and noise in these image data will bias the measurement \citep[e.g.][]{Massey2013,Viola2014}. If this bias is not accounted for, it can lead to incorrect determination of the intrinsic alignment signal \citep[Sect.  3.3.5]{Singh}. Fortunately, the bias can be characterized using images of simulated galaxies for which the input parameters are known. Ultimately, the precision in the bias measurement needs to be better than the statistical error on the alignment signal. Note that lowering the amplitude of the bias is not the most crucial task, but rather showing its robustness, i.e. how sensitive the bias is to changes of galaxy properties \citep{Hoekstra2017}. 

\begin{figure}
	\resizebox{\hsize}{!}{\includegraphics{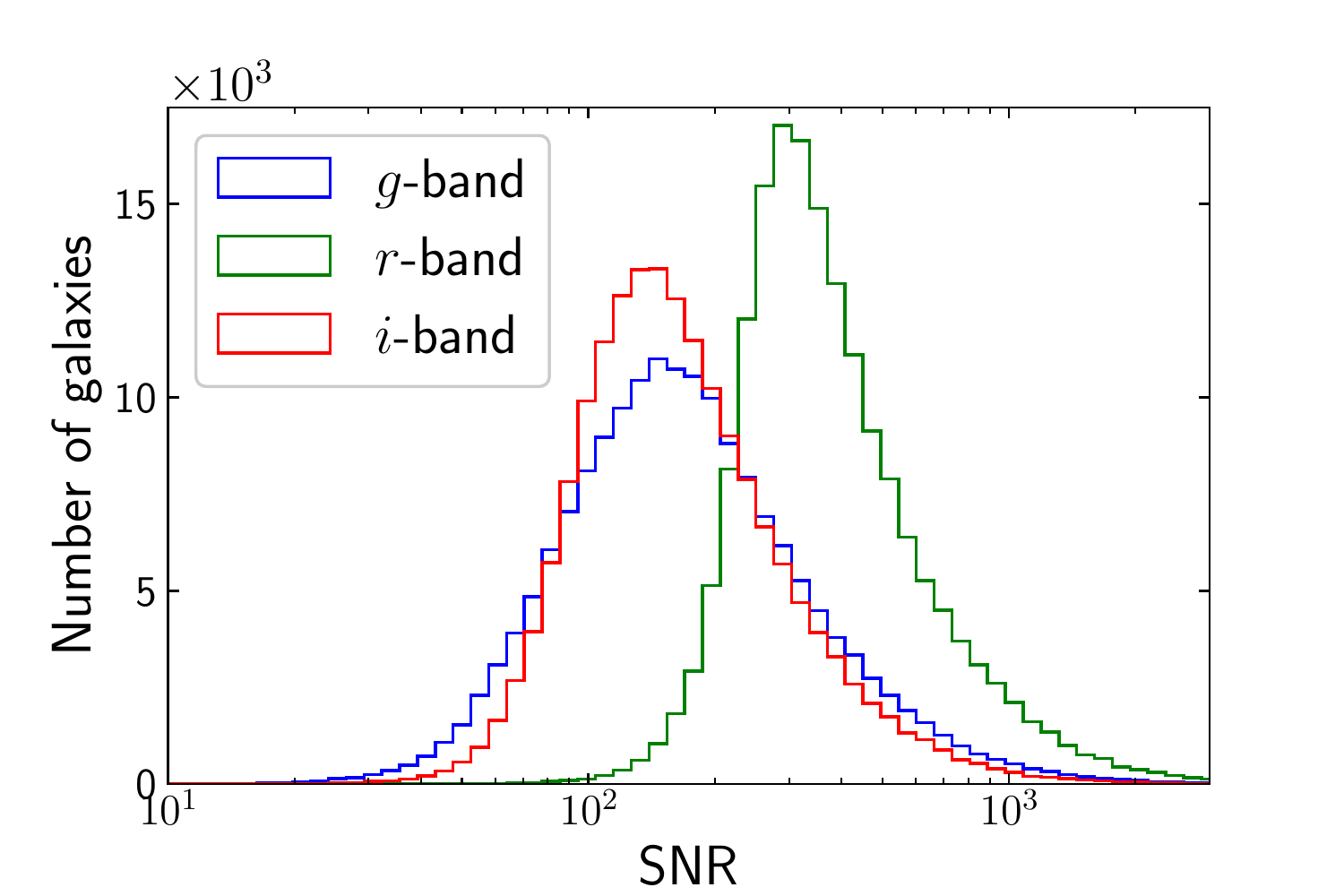}}
	\caption{Signal-to-Noise ratio distributions for GAMA galaxies in the KiDS $g$, $r$ and $i$-band images shown in green, red and blue, respectively. The SNR has been calculated using the weight function as in Eq. \eqref{eq:weight_function}.}
	\label{fig:KiDS_SNR}
\end{figure}

The performance of shape measurement methods depends on the input parameters of the image simulations. The bias is a function of size, ellipticity and SNR of the galaxies in the sample (but also depends on other properties of the simulated images, as discussed in \citealt{Hoekstra2017}). Consequently, the image simulations need to be as realistic as possible, representing closely the real data and galaxy properties. This requirement is not very strict in our case because the GAMA galaxies are very large, bright, with high SNR (Fig. \ref{fig:KiDS_SNR}), the PSF size is generally much smaller than the galaxies (see Fig. \ref{fig:R0_histograms}) and noise is not expected to give rise to very large biases for a reasonable weight function. With this in mind, the PSF used in the simulations is a Gaussian with FWHM equal to the mean FWHM of the KiDS images for the respective broad band filter. In Sect. \ref{sec:simulations-anisotropies} we explore the effect of shearing this PSF with an ellipticity equal to the PSF in the KiDS images.

We mimic our galaxy sample by using the S\'ersic photometry catalogue described in \cite{Kelvin2012} and the SNR measured from KiDS image data (Fig. \ref{fig:KiDS_SNR}). The catalogue is a single-S\'ersic fit produced with \textsc{Sigma}, a wrapper around several astronomical codes such as \textsc{SExtractor} \citep{Sextractor} and \textsc{GalFit} \citep{GalFit}. The fit is done for 167600 galaxies in the GAMA equatorial fields for each of the bands ugrizYJHK, using images from SDSS DR7 \citep{SDSSDR7} and UKIDSS-LAS \citep{UKIDSS}, with morphological properties extracted for every galaxy, such as S\'ersic index and half-light radius. The morphological parameters of the $r$-band images are used to produce image simulations representative of our galaxy sample \citep[see][Fig. 15]{Kelvin2012}. We require the S\'ersic fits for these galaxies to pass certain quality controls using the flags present in the catalogue. Specifically we only consider galaxies that have {\tt GAL\_QFLAG}, {\tt GAL\_GHFLAG} and {\tt GAL\_CHFLAG} equal to 0 in all $gri$ bands, which means that the final fit, global fitting history and component fitting history were not problematic. We also remove stars present in the catalogue by applying a redshift cut at $z>0.002$. Finally, we only keep galaxies for which a SNR measurement was possible in the KiDS imaging data for all three bands. This leads to a sample of 101209 galaxies from this catalogue.

We use \textsc{GalSim} \citep{GalSim}, a widely used Python package developed for the GREAT3 challenge {\citep{GREAT3}, to generate image simulations. The galaxy model we adopt is a S\'{e}rsic profile, motivated by the S\'ersic photometry catalogue, and the surface brightness of the profile is given by 

\begin{equation}
I(r)\propto \exp\left[-\beta_{n_{\rm s}}\left(\left(\frac{r}{r_{\rm e}}\right)^{1/n_{\rm s}}-1\right)\right]\,,
\end{equation}
where $n_{\rm s}$ is the S\'{e}rsic index, $r_{\rm e}$ is the half-light radius and $\beta_{n_{\rm s}}\approx2n_{\rm s}-0.324$ \citep{Capaciolli1989}. The galaxy profile is truncated at a radius of $4.5\cdot r_{\rm e}$. The galaxy is then sheared (in order to give each galaxy an intrinsic ellipticity) based on the ellipticity and position angle from the S\'ersic photometry catalogue, from which $n_{\rm s}$ and $r_{\rm e}$ are also obtained. These parameters are obtained from the $r-$band columns of the S\'ersic photometry catalogue and the same ones are used in the $g$ and $i$-band simulations.

The morphological parameters on $g$ and $i$-band images can be systematically different from the ones in $r$-band. We choose to fix them between simulations in order to quantify the bias in the shape measurement solely due to different image depth and quality. Moreover, we have checked that the measured bias does not depend strongly on these morphological parameters, so a slight systematic change in them will not affect our quoted biases significantly.

We only use galaxies with a S\'ersic index $0.3<n_{\rm s}<6.2$ (which holds for 96\% of the galaxies in the catalogue) as \textsc{GalSim} suffers from severe numerical problems for $n_{\rm s}<0.3$ (1\% of the galaxies) and for $n_{\rm s}>6.2$ (3\% of the galaxies) the shearing is not accurate \citep{GalSim}. The bias of the shape measurement depends on the galaxy's S\'ersic index and, in the case of DEIMOS, it increases (in absolute value) with increasing $n_{\rm s}$. However, we observed that this increase is not very rapid, and since only 3\% of our galaxy sample has $n_{\rm s}>6.2$ we do not expect this cut to affect our bias calibration significantly. For every galaxy we simulate two images that are rotated by 90 degrees, in order to eliminate intrinsic shape noise \citep[e.g.][]{Ian}.

We model the ellipticity bias similarly to \cite{STEP1}, with a multiplicative and additive term, given by
\begin{equation}
\epsilon_i^{\rm{obs}}=(1+m_i)\epsilon_i^{\rm{true}}+c_i\,,
\label{eq:bias}
\end{equation}
where $i=\{1,2\}$, $m_i$ is the multiplicative bias and $c_i$ is the additive bias. Note that, unlike the definition in \cite{STEP1}, the bias here concerns the \emph{ellipticity} and not the shear. Generally, we do not expect the bias in the ellipticity measurement to be linear since the ellipticity is a bounded quantity between 0 and 1, which is typically not small enough to ignore higher order biases \citep[e.g.][]{Miller2013, Pujol2017}. This is the main reason the ellipticity bias is not used when calibrating shear measurements. However, given the high SNR of our galaxy sample and the de-weighting procedure of \textsc{DEIMOS}, we show in Sect. \ref{sec:simulations-weightFunction} that this is very close to true for our shape measurements.

In the end, we need to define a postage stamp which will be used to measure the moments of each galaxy. The integrals involved in moments calculation theoretically extend from $-\infty$ to $+\infty$, and the postage stamp should be large enough to approximate this integral with enough accuracy. To determine how large the postage stamps should be, we calculate the integrand of the 6th order moment for a sheared S\'ersic profile (which as we will see in Sect. \ref{sec:simulations-weightFunction} is the highest order required for the shape measurement). We set the S\'ersic index to $6$ as a worst-case scenario, where the galaxy's profile extends far from the centroid. The postage stamp is then chosen by equating the 6th moment integrand with the noise level in the image, meaning that extending the postage stamp beyond this point will include noise dominated parts of the image and is therefore not expected to contribute to the moments calculation. To avoid extreme cases we set a minimum and maximum size of 50$\times$50 and $300\times300$ pixels, respectively, for the postage stamp. We find that increasing the maximum postage stamp size does not improve the accuracy of the measured shapes. This procedure results in optimizing the postage stamp size for each galaxy individually, using its flux, half-light radius and rough ellipticity (i.e. the \texttt{Elongation} output of \texttt{SExtractor}), thus reducing the runtime of the shape measurements significantly, compared to using a fixed postage stamp for each galaxy.

\subsection{Choosing the weight function}
\label{sec:simulations-weightFunction}

The weight function described in Sect. \ref{sec:DEIMOS} is defined for every galaxy by the size $r_{\rm wf}$ in equation \eqref{eq:weight_function}. This expresses the standard deviation of the initial circular Gaussian weight function employed in the first iteration of the matching procedure, and is generally different for every galaxy, depending on its surface brightness profile. In order for our shape measurement to be consistent between the three broad band filters, we determine $r_{\rm wf}$ in the $r$-band images, and use this value for the shape measurement in the other two filters. To choose an optimal value for this we investigated a range of possible choices for which we calculated the bias. This range is based on the radius of the circularised isophote of each galaxy, $r_{\rm iso}$, which is related to the area of the galaxy's isophote $A_{\rm iso}$ through
\begin{equation}
r_{\rm iso}=\sqrt{A_{\rm iso}/\pi}\,.
\label{eq:isophotal radius}
\end{equation}
$A_{\rm iso}$ is calculated using the \texttt{ISOAREA\_IMAGE} parameter from  \textsc{SExtractor}\footnote{The isophote is measured at 3$\sigma$ above the background noise RMS.}.

Another choice we need to make before measuring galaxy shapes is the maximum Taylor expansion order used for de-weighting $n_w$. The larger this number, the higher the order of moments used to correct for the weight function. Note that this number is even, since we have picked a symmetric weight function. To explore this, we picked $r_{\rm wf}=r_{\rm iso}$  and calculated the bias as a function of the input ellipticity for different values of $n_w$. We used equation \eqref{eq:bias} and performed a simple linear regression analysis for variables $m_i$ and $c_i$. 

The results are presented in Fig. \ref{fig:nwcheck}, where we show the mean multiplicative bias\footnote{Additive bias arises mostly from imperfect PSF modelling (or more generally from spurious ellipticity correlations) and is consistent with zero at the 3-sigma level for all our measurements, unless stated otherwise.} as a function of the input ellipticity of the galaxies in our simulation for four values of $n_w$. These results are for simulations of GAMA galaxies observed in the KiDS $r$-band filter. When no correction is applied ($n_w=0$ in Fig. \ref{fig:nwcheck}) the bias increases up to 5\% as the input ellipticity becomes large. Applying the simplest correction ($n_w=2$) already improves the shape measurement significantly. We can see that for small ellipticities the bias is around -2\% rising up to 1.5\% as we move to large input ellipticities. Further correction reduces this rising trend significantly, as seen in Fig. \ref{fig:nwcheck} for $n_w=4$. Including even higher order corrections to improve the de-weighting comes with a price. Besides increased computational time, the dispersion in the measured shapes becomes larger (i.e. the estimated ellipticities become more noisy) as shown in Fig. 2 of \cite{DEIMOS}. This is also seen in Fig. \ref{fig:nwcheck}, where the errors and the noise bias become larger as  $n_w$ increases. We adopt $n_w=4$, given the fairly constant response of the bias to input ellipticity (compared to $n_w=2$) and the relatively low effect of noise (compared to $n_w=6$).

\begin{figure}
	\resizebox{\hsize}{!}{\includegraphics{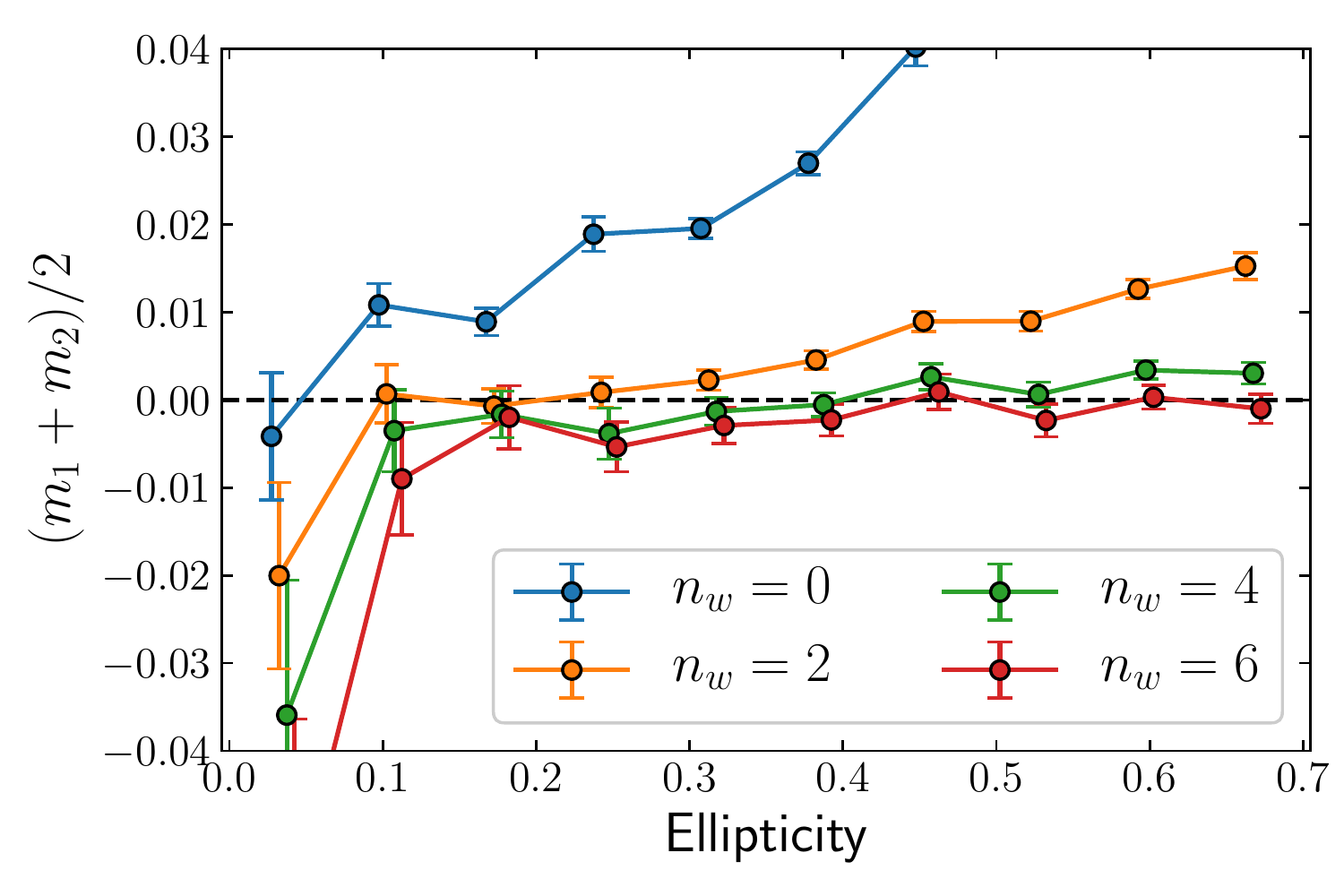}}
	\caption{Mean multiplicative bias as a function of ellipticity for different values of the maximum Taylor expansion order $n_w$. Points are horizontally displaced for clarity.}
	\label{fig:nwcheck}
\end{figure}

Having determined the maximum Taylor expansion order to use, we need to find the optimal value for $r_{\rm wf}$, the size of the initial circular weight function. We do so by calculating the multiplicative bias as a function of $r_{\rm wf}/r_{\rm iso}$. The results are shown in Fig. \ref{fig:wfcheck}. The bias seems to be fairly constant for an initial weight function between $0.75$ and $1.5$ times the circularized isophote of each galaxy. A larger weight function results in an increase in bias as well as a decrease in the accuracy with which the bias can be measured. This is caused partly due to the shape measurements being more noisy but also due to loss of statistical power, as more and more shape measurements are flagged as problematic (since larger weight function results in including more noise in the measurements). For a very small weight function the bias is positive, and we see a point where the bias must be zero. However, this does not mean that the bias is better calibrated because the noise bias has been traded off with model bias. In the end, we choose to use an initial weight function of size $r_{\rm wf}=r_{\rm iso}$, around which the bias remains fairly constant.

\begin{figure}
	\resizebox{\hsize}{!}{\includegraphics{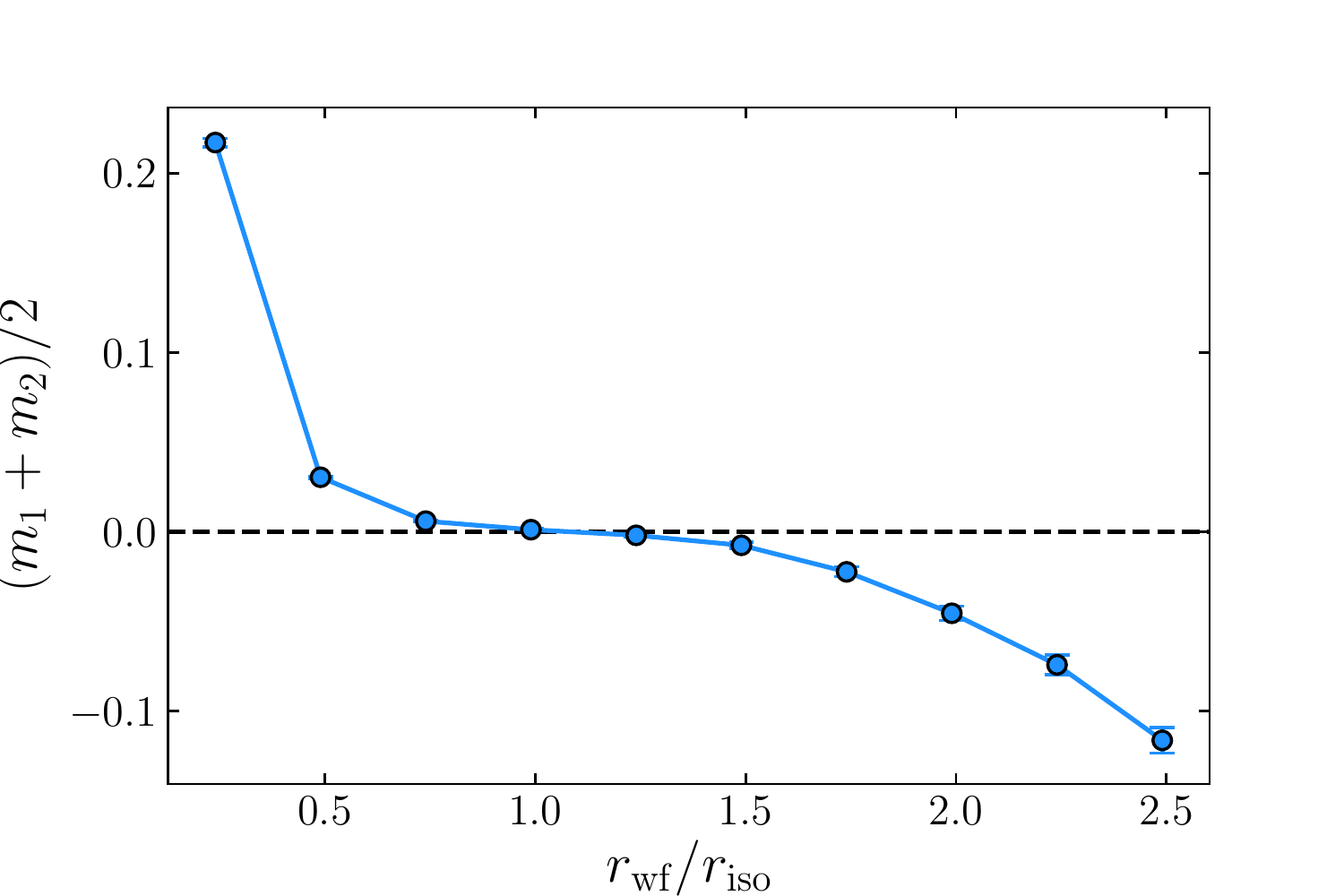}}
	\caption{Mean multiplicative bias as a function of the size of the initial weight function used in our simulations.}
	\label{fig:wfcheck}
\end{figure}

\subsection{Masking and PSF anisotropies}
\label{sec:simulations-anisotropies}

Galaxies observed in imaging data are generally not isolated but have neighbouring galaxies. These neighbours need to be masked in order to measure the shape of the galaxy robustly. The masking is done using the segmentation map produced by \textsc{SExtractor}, which identifies the pixels associated with every source in the image. The pixels \emph{not} associated with the galaxy, whose shape is measured, are replaced with random Gaussian noise with variance equal to the background noise RMS. 

However, this masking affects the integration required to calculate the galaxy's moments, since essentially there are certain values of the $x,y$ Cartesian coordinates over which the galaxy's flux is replaced by the expected noise level. Consequently, the masking of neighbouring galaxies will introduce an extra source of bias in the shape measurements. Here we aim to quantify this bias using our image simulations. For every galaxy simulated, we also create a cut-out of the segmentation map obtained from the KiDS imaging data for this particular galaxy, and measure the galaxy's shape using this segmentation map. We observe a general shift of the bias towards lower values, similar to the effect of noise bias. For example, the mean ellipticity $m$-bias in the $r$-band simulation shifts from a value of 0.12 \% to -0.37 \%. 

In addition to masking, anisotropies in the PSF are also expected to impact shape measurements to some degree. With DEIMOS, however, this is not expected to be important since the PSF convolution is treated in an exact analytical way.
In addition, the galaxies in our sample are much larger than the PSF (Fig. \ref{fig:R0_histograms}) and their shape determination is not expected to be significantly affected by this. We test these with our simulations by considering elliptical Gaussian PSF, arbitrarily oriented, with its ellipticity equal to the mean PSF ellipticity of KiDS images, which is approximately $\epsilon_{\rm PSF}=0.05$ in all three $gri$ filters \citep{deJong2017}. A small shift in the recovered $m$-bias of the order of 0.01\% was observed, which is negligible compared to the statistical uncertainty with which the bias is determined (i.e. the statistical error in the linear regression analysis). 

\begin{table*}
	\caption{Multiplicative and additive bias obtained from our image simulations in the three $gri$ broad-band filters. }
	\label{tab:biases}
	\centering
	\begin{tabular}{l c c c c}
		\hline\hline
		Simulation & $m_1$ & $m_2$ & $c_1$ & $c_2$ \\
		\hline
		$g$-band & $-0.0073\pm0.0016$ & $-0.0045\pm0.0017$  & $(24.9\pm34.0)\times10^{-5}$ & $(-8.3\pm35.2)\times10^{-5}$ \\
		$r$-band & $-0.0040\pm0.0008$ & $-0.0035 \pm 0.0008$ & $(-4.3\pm18.2)\times10^{-5}$ & $(-11.0\pm18.1)\times10^{-5}$\\
		$i$-band & $-0.0084\pm0.0017$ & $-0.0083\pm0.0016$ & $(-0.5\pm35.2)\times10^{-5}$ & $(-10.2\pm34.6)\times10^{-5}$ \\
		\hline
	\end{tabular}
\end{table*}

\subsection{Bias of the ellipticity}
\label{sec:simulations-ebias}

We now quantify the shape measurement bias on the ellipticity of our galaxy sample. Choosing $r_{\rm wf}=r_{\rm iso}$ and $n_w=4$, we measure ellipticities $\epsilon_i^{\rm obs}$ and calculate the bias for simulated GAMA galaxies in $gri$ images, using the SNR measured from KiDS $g$, $r$ and $i$-bands. The initial weight function $r_{\rm wf}=r_{\rm iso}^r$ is determined from simulated $r$-band images (the superscript defines the band used to determine $r_{\rm iso}$) and is then also used for the $g$ and $i$-band images. We choose to use the $r$-band measured isophote since $r-$band images are of higher quality and we wanted to measure the shape of the same parts of the galaxies by using the same weight function on all three bands. The actual $r_{\rm iso}^g$ and $r_{\rm iso}^i$ values are not very different than $r_{\rm iso}^r$, and as we can see from Fig. \ref{fig:wfcheck} the bias is fairly constant around $r_{\rm wf}=r_{\rm iso}$, therefore this choice will not result in significantly larger biases in the $g$ and $i$-band. However, galaxies imaged in the $g$ and $i$-band filters have lower SNR compared to the $r$-band images and therefore we expect a larger noise bias on the $g$ and then $i$-band shapes.

In Table \ref{tab:biases} we present the bias of measured shapes on simulated galaxies for the three broad band filters.  As a
sanity check, we note that $m_1 = m_2$ , within the error bars,
for each set of simulated galaxies. In addition, additive biases are consistent with zero. The lowest bias is obtained for $r$-band image simulations, as expected, followed by $i$ and $g$-bands.

\section{Results}
\label{sec:results}

Having calibrated the shape measurement method against realistic image simulations, we present our results in this section. Shapes of galaxies in $g,r$ and $i$-band images were measured and the final sample consists of galaxies for which the shape was successfully measured in all three filters, which is the case for 89.7\% of the initial galaxy sample. We have checked that the galaxies that were rejected follow the same redshift distribution as the whole galaxy sample. We first examine the distributions of ellipticity and size for our galaxy sample in the three filters. Then, we investigate differences in the intrinsic alignment signal measured in the three filters, and try to understand the source of the observed difference, by splitting the galaxy sample into further sub-samples based on colour, redshift and central/satellite galaxies. 

\subsection{Ellipticity and size distributions}
\label{sec:edistr}

The ellipticity distributions (after applying the bias correction) for all the galaxies in our sample, as measured from the three filters, are shown in Fig. \ref{fig:edistr}. 
\begin{figure}
	\resizebox{\hsize}{!}{\includegraphics{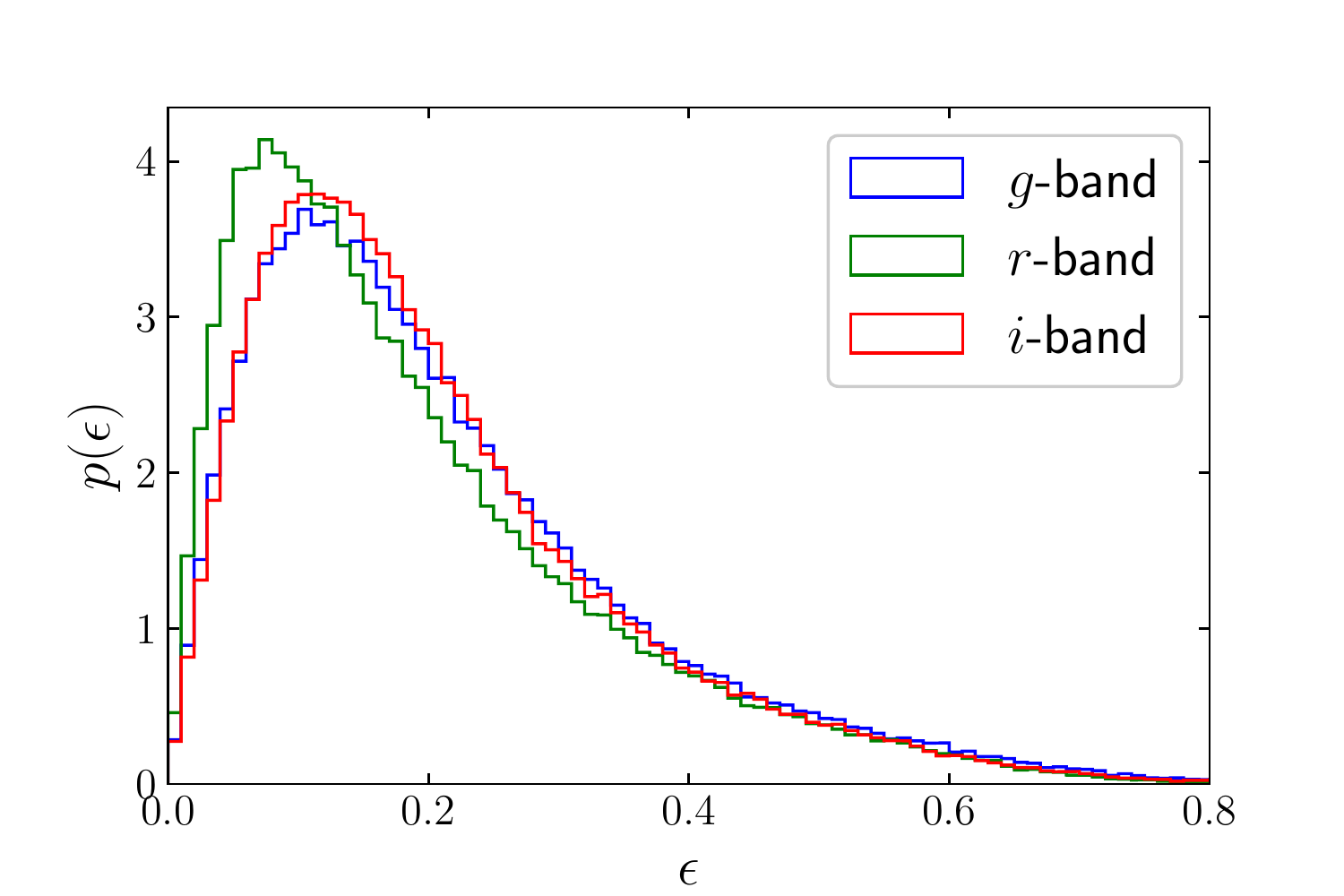}}
	\caption{ Distribution of ellipticity of GAMA galaxies as measured from the three broad band $gri$ KiDS images. The qualitative difference observed between the $r$-band and the ellipticities in the other filters is found to be caused due to higher noise in the $g$ and $i$-band ellipticities.}
	\label{fig:edistr}
\end{figure}
The shapes from the $g$ and $i$-band images have similar ellipticity distributions while the ellipticity measured in the $r$-band follows a distribution that peaks at lower $\epsilon$ and drops more quickly as $\epsilon$ increases. This behaviour is expected solely due to the non-linear nature of the ellipticity (i.e. a ratio of surface brightness moments). As shown in \citet{MelchiorViola}, noise causes the ellipticity distribution to be generally asymmetric and skewed, pushing its peak to higher values. As noise increases, this peak is pushed further away from its true value and we see this behaviour in our image simulations as well. Keeping all morphological parameters the same, the measured ellipticity distribution of our $g$ and $i$-band image simulations is pushed to slightly higher values than the higher-SNR $r$-band image simulations.

However, the ellipticity distributions in Fig. \ref{fig:edistr} are much more significantly different compared to what we see in our image simulations. We visually inspected galaxies with largely different ellipticities between $r$-band and $g$ or $i$, and discover that most of them where due to relatively bright concentrated circular bulges present in the $r$-band images. Therefore, these differences, while partly caused due to noise, are also attributed to morphological differences in the galaxy $g,r$ and $i$-band images. We also note that the ellipticity distributions where observed to be more discrepant in high redshift galaxies ($z>0.26$) compared to low redshift. This suggest that, at least partly, these differences could be caused by the fact that different parts of a galaxy's spectral energy distribution is observed in a given filter for galaxies at varying redshift. In Sect. \ref{sec:edistr} we discuss the possible contamination to the intrinsic alignment measurement from these different ellipticity distributions.

As mentioned in Sect. \ref{sec:Introduction}, galaxy colour gradients tend to make galaxies appear larger in blue than in red filters. Having measured quadrupole moments for galaxies in the three $gri$ broad bands, we are in a position to test this hypothesis. We quantify the size of a galaxy as
\begin{equation}
S_{\rm gal} = \sqrt{Q_{20}Q_{02}-Q_{11}^2}\scalebox{1.75}/Q_{00}\,,
\label{eq:sizes}
\end{equation}
where we use the deconvolved, de-weighted moments $Q_{ij}$. The comparison is presented in Fig. \ref{fig:sizes}, where the density map is shown for the sizes measured in the three filters. We can see that sizes measured in the $g$-band are generally larger than those measured in the $r$-band, in agreement with our expectations. The $r$ and $i$-band sizes seem to be very similar, hinting, however, towards slightly larger sizes for galaxies in the $r$-band. Note that the weight function used for the shape measurements have been defined in the $r$-band images, and the differences could potentially be larger than what is seen in Fig. \ref{fig:sizes}.

\begin{figure}
	\resizebox{\hsize}{!}{\includegraphics{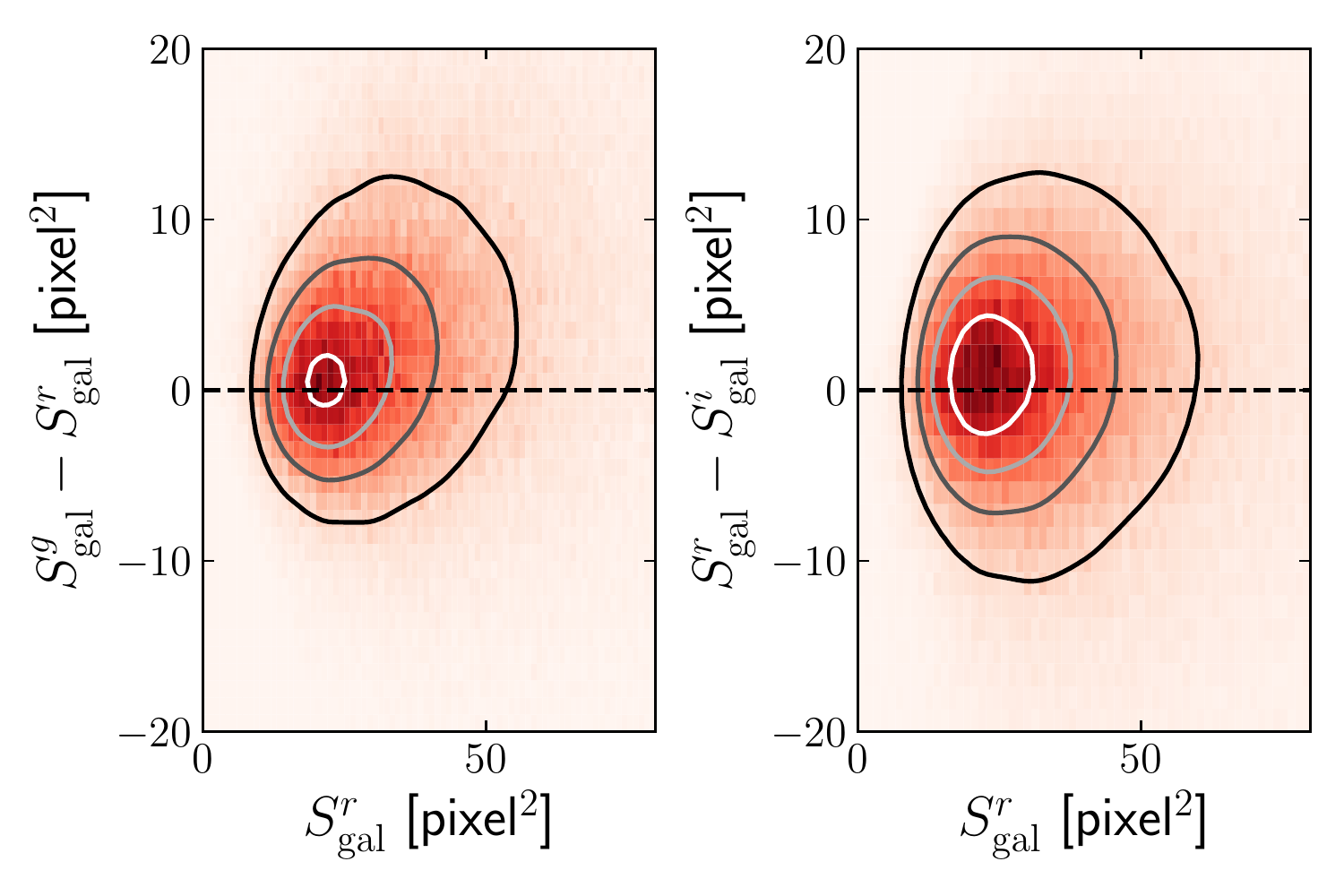}}
	\caption{Comparison of galaxy sizes, given from Eq. \eqref{eq:sizes}, between $g$ and $r$-band sizes (left) and $r$ and $i$-band (right). The density map is shown, with denser regions appearing redder. Equally spaced contours are overlaid for clarity.}
	\label{fig:sizes}
\end{figure}

\subsection{Intrinsic alignment measurement methodology}
\label{sec:IAdiff}

Intrinsic alignments are commonly quantified by the projected correlation function between galaxy ellipticity and galaxy density, 
\begin{equation}
w_{g+}(\mathbf{r}_p)=\int_{-\Pi_{\max}}^{+\Pi_{\max}} \xi_{g+}(\mathbf{r}_p, \Pi)\mathrm{d}\Pi\,,
\label{eq:wg+theory}
\end{equation}
where $\xi_{g+}$ is the three-dimensional correlation function. The line-of-sight distance between the galaxy pairs is $\Pi$ and $\mathbf{r}_p$ is the transverse separation of these galaxies. We measure $\xi_{g+}$ using the modified Landy-Szalay estimator\footnote{For a discussion on the choice of the estimator used see \citet{Johnston}.} \citep{LandySzalay, edo2017}, given by
\begin{equation}
\xi_{g+}(\mathbf{r}_p, \Pi)=\frac{S_+D}{D_{\rm S}D}-\frac{S_+R}{D_{\rm S} R}\,,
\label{eq:xig+}
\end{equation}
In the above equation capital letters define a particular galaxy sample, with $S_+$ for a sample of galaxy shapes, $D$ for for galaxy density and $R$ containing random points in the survey area. $D_S$ is the density field traced by the sample of galaxy shapes. $D_SD$ and $D_SR$ are the (normalised) number of pairs in the respective catalogues \citep[see Eq. 16 in][for a detailed description]{Kirk2015}. The alignment is essentially quantified by
\begin{equation}
S_+D = \sum_{i\neq j}\epsilon_+ (i|j)\,,
\label{eq:S+D}
\end{equation}
where galaxy $i$ is taken from the shape sample and $j$ from the density sample (or the random sample, for $S_+R$). The tangential ellipticity component $\epsilon_+(i|j)$, defined between galaxies $i$ and $j$, projects the ellipticity along the vector connecting the two galaxies, and is defined by
\begin{equation}
\epsilon_+=\epsilon_1 \cos 2\theta_p+\epsilon_2\sin 2\theta_p\,,
\label{eq:eplus}
\end{equation}
where $\theta_p$ is the angle between the x-axis of the coordinate system and the line connecting the galaxy pair. The ellipticity components $e_1, e_2$ are defined for galaxy $i$ with respect to the x-axis of the coordinate system. A positive/negative value of $\epsilon_+$ implies radial/tangential alignment. Rotating by 45 degrees, we can define the cross ellipticity component as
\begin{equation}
\epsilon_\times = \epsilon_1\sin 2\theta_p-\epsilon_2\cos 2\theta_p\,.
\label{eq:ecross}
\end{equation}

The ellipticities are corrected for multiplicative bias according to Table \ref{tab:biases}. We note here that the intrinsic alignment signal is measured using ellipticities and not shear, with the latter usually preferred when quantifying the contamination of IA to weak lensing measurements. For the random sample $R$ we use random catalogues specifically designed for GAMA \citep{Farrow}, down-sampled to contain roughly 100 times more galaxies than the number of galaxies in the shape sample, which we confirmed to be sufficiently large to produce consistent random signals. In an analogous way, $w_{g\times}$ can be measured, which is expected to be zero and can serve as a test for systematic errors. 

The correlation function in equation \eqref{eq:wg+theory} can be connected to analytical predictions for intrinsic alignments. For details on modelling the intrinsic alignment signal, as well as a discussion on the contamination of this signal to future cosmic shear surveys, we refer the reader to \citet{Johnston}. In this work we will focus only on potential differences in the measured intrinsic alignment signal between the different broad band filters. To measure projected correlation functions, we integrate over $-60\le \Pi \le 60$ Mpc$/h$ (in bins of $\Delta\Pi=4$ Mpc$/h$) and then consider 11 bins of transverse separation $r_p$, logarithmically spaced between $0.1$ and $60$ Mpc$/h$. 

We obtain the covariance matrix of the data, along with error bars, accounting for shape measurement noise. We do so by creating 100 realisations of the shape catalogues, adding a random position angle to each galaxy (making sure the same angle is added to the same galaxies in all three filters for each realisation) while keeping their original ellipticity modulus. We expect sample variance to not play an important role since we are measuring the difference in the alignment signal between broad band filters using exactly the same galaxies. Potential differences in the alignment signal are not expected to arise from large scale structure (and not affected strongly from clustering), but should be attributed to a rather local mechanism. Even though we cannot guarantee that the sample variance is zero, we are confident any effect of it in the error estimation is sub-dominant.

\subsection{Intrinsic alignment differences in the $gri$ filters}

We now quantify the difference in the intrinsic alignment signal measured among the three $gri$ broad bands. This is expected to be zero if the correlation of galaxy shapes and positions is not systematically different between the three bands. Therefore, measuring a non-zero difference in the correlation functions would suggest that the intrinsic alignment signal is different for observations at different wavelengths. We note that, for each galaxy, the same weight function has been applied to all three broad band images, and only galaxies with reliable shapes in all three images are used for the shape sample. This ensures that the exact same galaxies are used to calculate the intrinsic alignment signal in each filter, and the weight function is fixed on the same physical size for each galaxy. Several tests for systematics were performed, which did not reveal any problems with the analysis (see Appendix \ref{sec:systematics})

We calculate $\Delta w_{g+}=w_{g+}^{\rm band1}-w_{g+}^{\rm band2}$, among the three bands ($g-r$ and $r-i$)\footnote{We check that the $\Delta w_{g+}$ for the $g-i$ filter combination is roughly the sum of the other two combinations and is therefore omitted.} as a function of the transverse separation of the galaxy pairs, $r_p$. We use all galaxies with reliable shapes in unmasked regions for the shape population and all the galaxies for the density population. We perform a redshift cut $0.01<z<0.5$ which matches the redshift range over which the random catalogue was generated \citep{Farrow}. The results are shown in Fig. \ref{fig:wgplus}. The difference in the correlation functions is non-zero for small galaxy separations, $r_p\lesssim2$ Mpc$/h$, whereas, on the largest scales the difference in the signals is consistent with zero. By examining the individual IA signal for the $w_{g+}^{\rm g,r,i}$ we see that they are all positive (indicative of radial alignments, where the semi-major axis of galaxies points towards each other). Conclusively, the signal is lower in the $r$ band, compared to $g$ and $i$.

\begin{figure}
	\resizebox{\hsize}{!}{\includegraphics{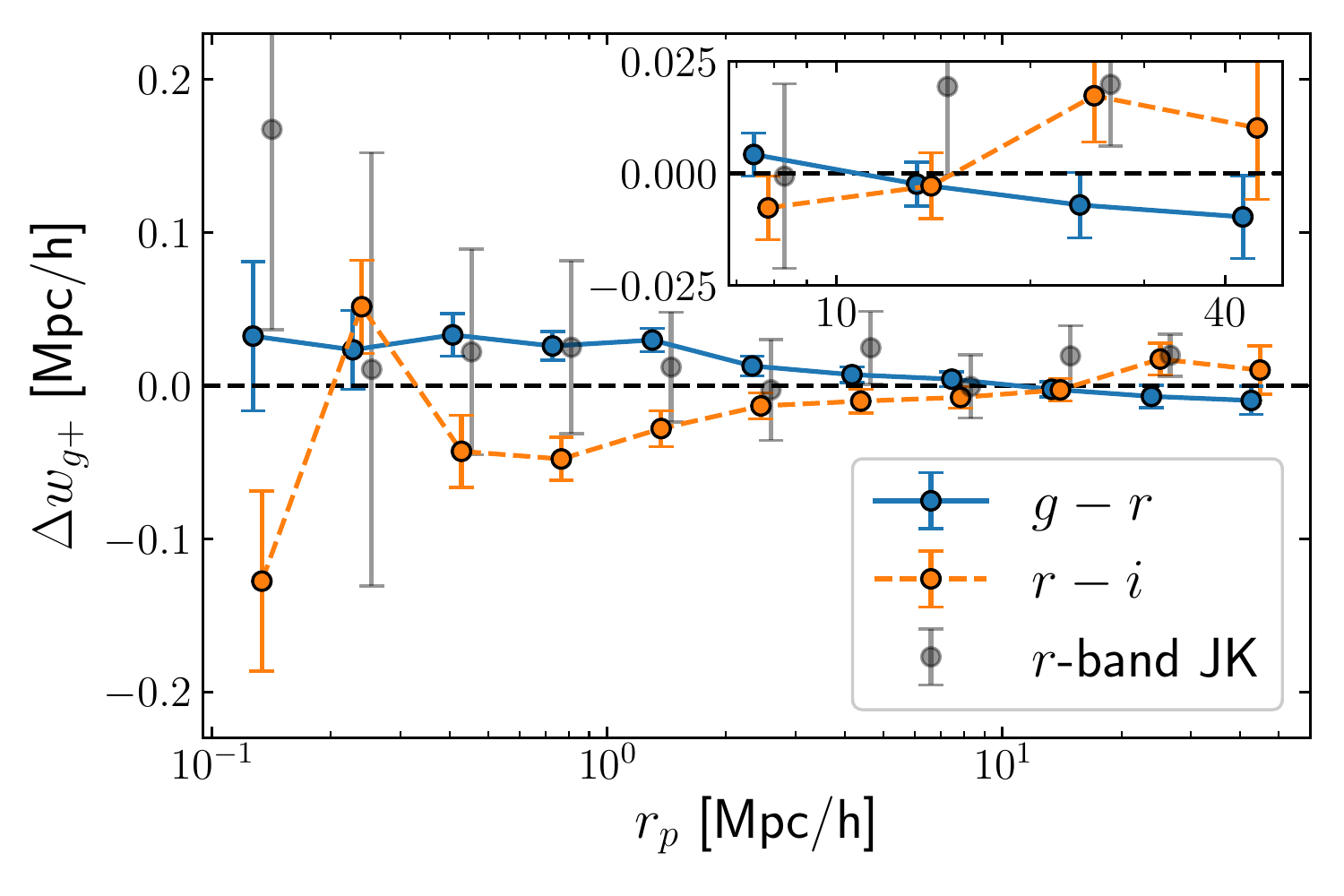}}
	\caption{Difference in projected density-shape correlation function between the $g-r$ and $r-i$ filters (blue and orange points respectively). The x-axis shows the transverse separation of galaxy pairs. For reference, the signal measured in $r$-band is shown with grey points, with errors obtained from 3D jackknife resampling. The in-line figure is a zoom-in on the last four data points, in the highest separations. Points are horizontally displaced for clarity.}
	\label{fig:wgplus}
\end{figure}

As a reference, we show the intrinsic alignment signal measured in the $r$-band for scales up to $\sim30$ Mpc/$h$. The error bars are estimated using a jackknife technique to obtain the covariance matrix, cutting the survey area in 3D cubes to achieve large enough number of jackknife samples. The GAMA survey geometry limits us to using scales only up to $\sim30$ Mpc/$h$, since covariance on larger scales cannot be captured by the jackknife configuration. For a detailed analysis on the jackknife error estimation, as well as the $r$-band signal itself, we refer the reader to \citet{Johnston}.

We see that, at large scales, the difference is generally smaller than the $r$-band signal. At smaller scales, below 2 Mpc$/h$, the difference is larger, even though the $r$-band signal is generally noisy at these scales\footnote{We note here that the error bars on $\Delta w_{g+}$ are significantly smaller than the ones in the individual signals. We are able to calculate such small error bars because we use exactly the same galaxies to calculate $\Delta w_{g+}$ and are only left with the errors due to shape measurement noise, which are generally very small.}. Interestingly, the $\Delta w_{g+}$ values for these scales are comparable to the value of $w_{g+}$ of the $r$-band measurement on large scales. Through this comparison, we conclude that the difference observed in alignment with wavelength is large enough that it cannot be neglected.

Since the data points are correlated, we use the full shape-noise covariance matrix and a $\chi^2$ analysis to assess the significance of the signals. We test the 11 data points against a null-signal hypothesis and quote the $p$-values for 95\% confidence level. The difference in alignment signal is significantly non-zero, both in $g-r$ and $r-i$ measurements, with $p$-values of 0.02015 and 0.00048, respectively. We also restrict the analysis to the largest scales, beyond $r_p>6$ Mpc$/h$, which includes the last four data points shown zoomed in, in the in-line figure of Fig. \ref{fig:wgplus}. For these large scales the difference in signals is consistent with zero, with $p$-values equal to 0.56 and 0.16 for the $g-r$ and $r-i$ difference, respectively.

\subsection{Tracing the origin of the difference}

In order to investigate the source of the wavelength dependence of the intrinsic alignment signal, evident in Fig. \ref{fig:wgplus}, we calculate the IA signal splitting the galaxy sample into sub-samples. Using the stellar masses catalogue \texttt{StellarMassesLambdarv20} from the GAMA DMU \citep{StellarMasses} we acquire colour information for our galaxy sample and split in intrinsically red and blue galaxy population (using the rest-frame $g-i$ colours of the stars in the galaxy, \texttt{gminusi\_stars}>0.75 and <0.75, respectively). We observe that $\Delta w_{g+}$, measured for these two populations, is non-zero in both cases but larger in amplitude for red galaxies. They are also seen at different scales: on smaller scales for the blue galaxy sub-sample and on larger scales for the red sub-sample (see Appendix \ref{sec:redshift_color}). This behaviour is expected according to the tidal alignment model and tidal torque model (for pressure and rotationally supported galaxies, respectively, see \citealt{Kiessling2015}), according to which alignments between blue galaxies manifest generally on small scales, while alignments in red galaxies can be observed further out, and generally reach higher amplitudes. For blue galaxies, any difference in the signal would be observed on smaller scales as well.

Moreover, we split the whole galaxy sample into low and high redshift sub-samples ($z<0.26$ and $z>0.26$, respectively). We observe a null $\Delta w_{g+}$ for high redshift galaxies, while the $\Delta w_{g+}$ for the low redshift sample looks very similar to what is seen in Fig. \ref{fig:wgplus} (see appendix \ref{sec:redshift_color}). While this would suggest an evolution of the intrinsic alignment difference with time, we find that the alignment signal does not vary greatly with redshift \citep[see e.g.][]{Johnston}. On the other hand, since our galaxy sample is flux limited, less luminous galaxies are observed in the lower redshift sample compared to the higher one. More interestingly, due to this fact, at lower redshift we observe more satellite galaxies than at high redshift. 

To investigate whether satellite galaxies can affect the difference in intrinsic alignment signal between bands, we use the Group catalogue of the GAMA DMU \cite{Groupcatalogue}. We find that the lower redshift sample contains more than twice as many satellite galaxies than the high redshift one. Furthermore, we split our sample of galaxies into satellite (column \texttt{RankBCG$>1$} in the Group catalogue) and central galaxies, where we take all the ``field'' galaxies (i.e. those not associated with a group) and all the brightest central galaxies (BCG) of groups as the central galaxies sample (\texttt{RankBCG$\le1$}). The reason for including the field galaxies is that we expect most of these to be the BCG of a galaxy group whose satellites are too faint to be observed.

\begin{figure*}
	\centering
	\includegraphics[width=17cm]{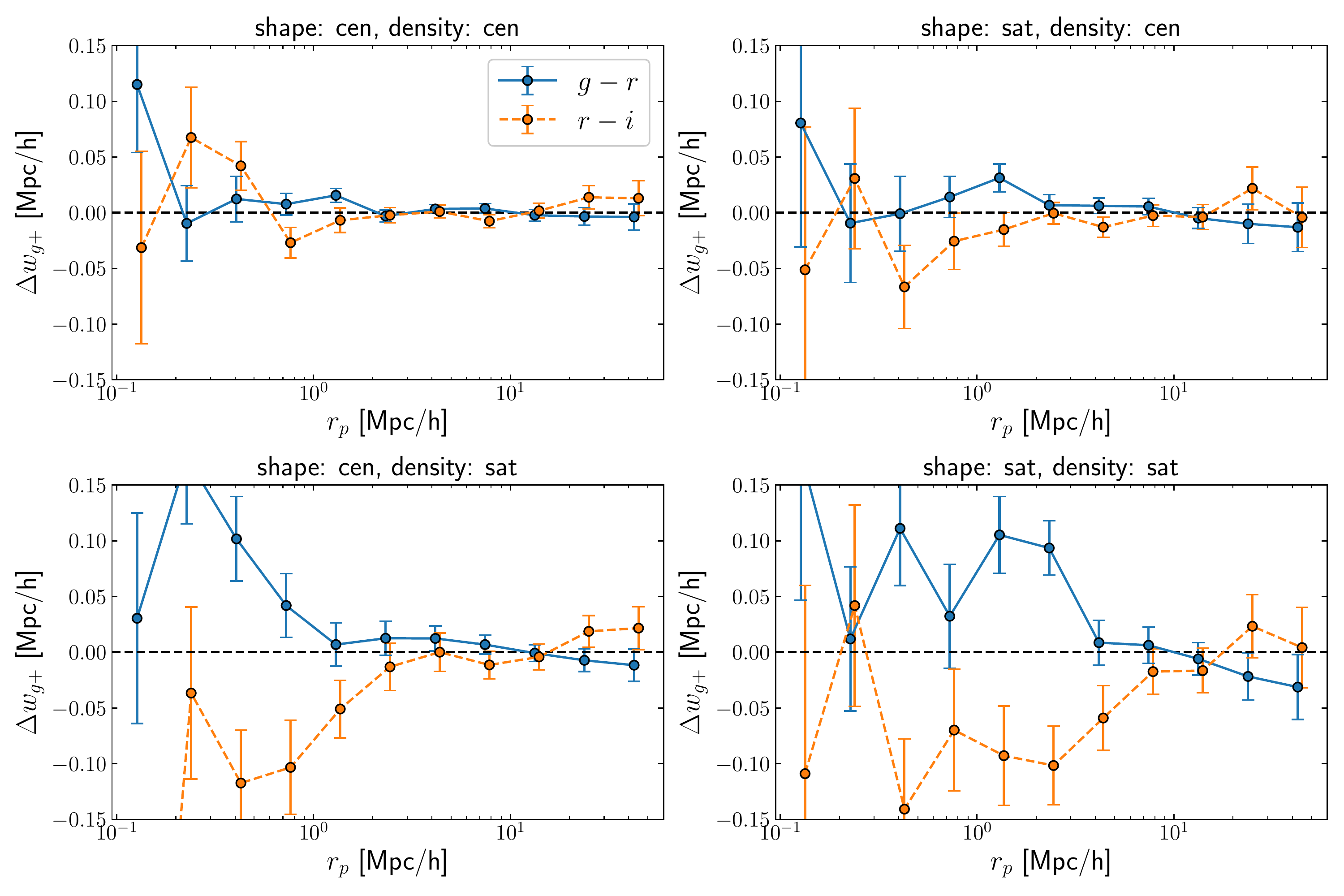}
	\caption{Difference in projected density-shape correlation function between the $g-r$ and $r-i$ filters (blue and orange points, respectively). The correlations were measured using the following shape/density samples: centrals vs centrals (top left), satellites vs centrals (top right), centrals vs satellites (bottom left) and satellites vs satellites (bottom right). The x-axis shows the transverse separation of galaxy pairs. Points are horizontally displaced for clarity.}
	\label{fig:wgpsat}
\end{figure*}

We correlate the shapes of central galaxies against the density of central galaxies and plot the difference in the obtained projected correlation function $\Delta w_{g+}$ in Fig. \ref{fig:wgpsat}. We find that the difference in alignment is consistent with zero between the three broad band filters for this sample of galaxies. The same conclusion is reached when correlating shapes of satellite galaxies against the density of central galaxies. Note that we do not restrict the correlation measurement to satellites and their corresponding group BCG, but rather correlate all identified satellites against all other galaxies in the sample. When the alignment signal is measured using shapes of central galaxies correlated with the density of satellite galaxies, the difference $\Delta w_{g+}$ between the three bands is significantly non-zero at scales below 1 Mpc$/h$, reaching values roughly twice as much as the ones seen in Fig. \ref{fig:wgplus}. As in Sect.  \ref{sec:IAdiff}, the IA signal $w_{g+}^{\rm g,r,i}$ is positive in each band, and hence the alignment signal in the $r$-band is the weakest. The same behaviour is observed when correlating shapes of satellite galaxies against position of satellites. In this case, the difference in the alignment signal extends further, up to around 3 Mpc$/h$. 

To further pinpoint the galaxy population responsible for the observed IA difference, we split the satellite galaxies into intrinsically red and blue, in the same way as for the whole population, described above. When blue satellites are considered, the intrinsic alignment difference has a very small amplitude and becomes noisy. When we correlate shapes and positions of red satellite galaxies, however, the IA difference is larger, up to twice the amplitude seen in Fig. \ref{fig:wgpsat}. This result can be seen in Fig. \ref{fig:wgpsatred}.

\begin{figure}
	\resizebox{\hsize}{!}{\includegraphics{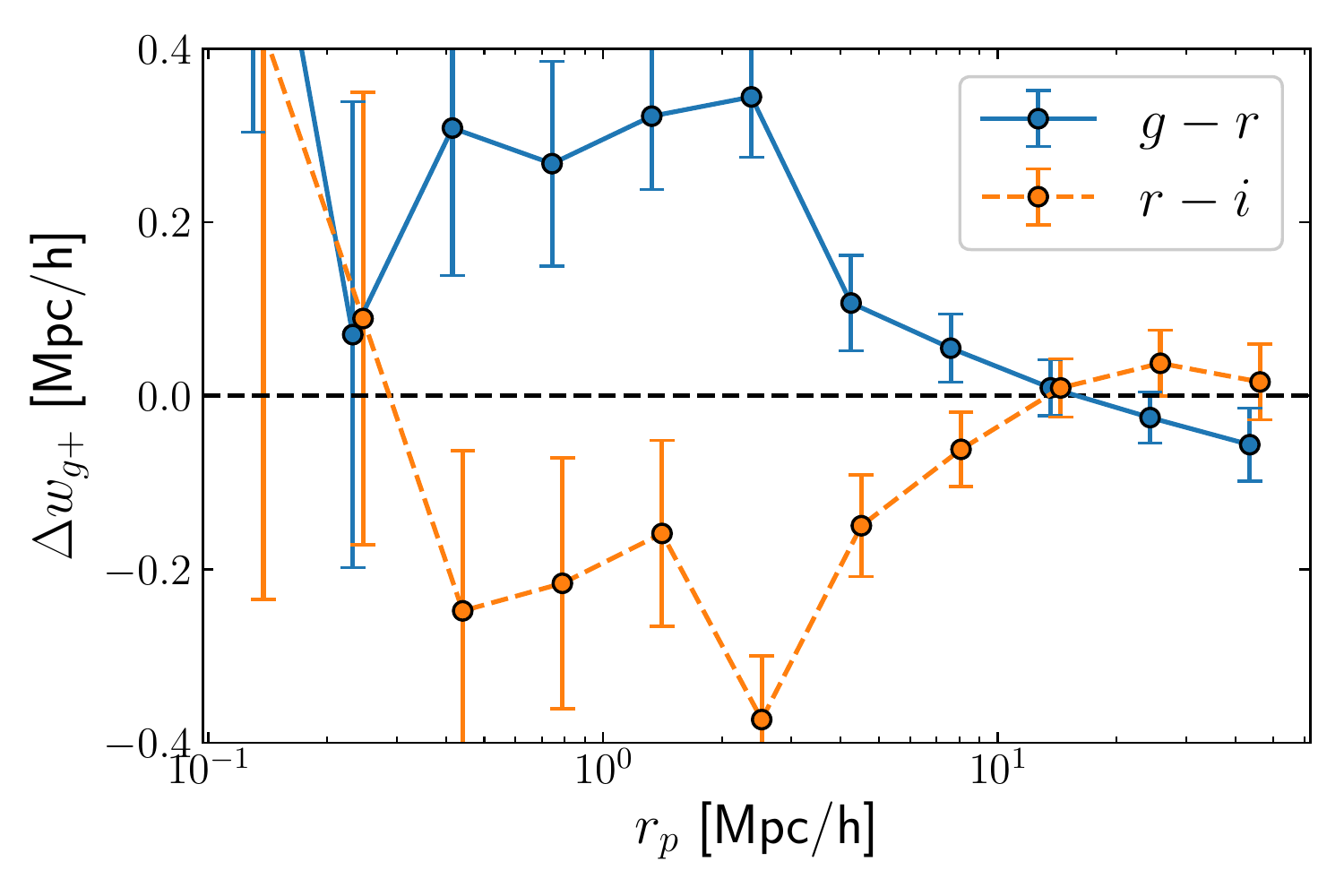}}
	\caption{Difference in projected density-shape correlation function between the $g-r$ and $r-i$ filters (blue and orange points, respectively) measured using red satellites as shape and density samples. The x-axis shows the transverse separation of galaxy pairs. Points are horizontally displaced for clarity.}
	\label{fig:wgpsatred}
\end{figure}

\subsection{Investigating ellipticity distribution differences}

As seen in Fig. \ref{fig:edistr}, the distributions in ellipticity for the $g,r$ and $i$ band shape measurements are not identical. Here, we investigate whether our results are a manifestation of this ellipticity difference. The first thing we looked at was the distribution of ellipticity for satellite and central galaxies. In their respective bands, these ellipticities were shown to be very similar. If the ellipticity distribution differences were to play a major role in our results, the $\Delta w_{g+}$ of central galaxies should be similar to that of satellite galaxies, which is clearly not the case, as seen in Fig. \ref{fig:wgpsat}. 

The $w_{g+}$ estimator described in \eqref{eq:wg+theory}-\eqref{eq:eplus} is essentially weighted by the absolute value of the galaxy's ellipticity $|\epsilon|$. To nullify the effect of the ellipticity distribution to the calculation of correlations, we calculate $w_{g+}$ using \emph{normalised} ellipticities for each galaxy, i.e. $\epsilon_i\mapsto \epsilon_i/|\epsilon|$. This ensures that the alignment is not weighted by the ellipticity of each galaxy. The results obtained using this ``pure'' alignment estimator are shown in Fig. \ref{fig:wgpnorm}. We see that the IA difference is still non-zero, the amplitude is now larger and the measurement has become more noisy. This is expected because more elliptical galaxies will show their alignment more clearly when weighted by their ellipticity, hence the noisier measurement of the normalized $\Delta w_{g+}$. On the other hand, nearly round galaxies will have a boosted signal when the estimator is divided by $|\epsilon|$, hence the larger amplitude. We conclude that the measured IA difference is not caused by the difference in ellipticity distributions of galaxies measured in the three broad band filters.

\begin{figure}
	\resizebox{\hsize}{!}{\includegraphics{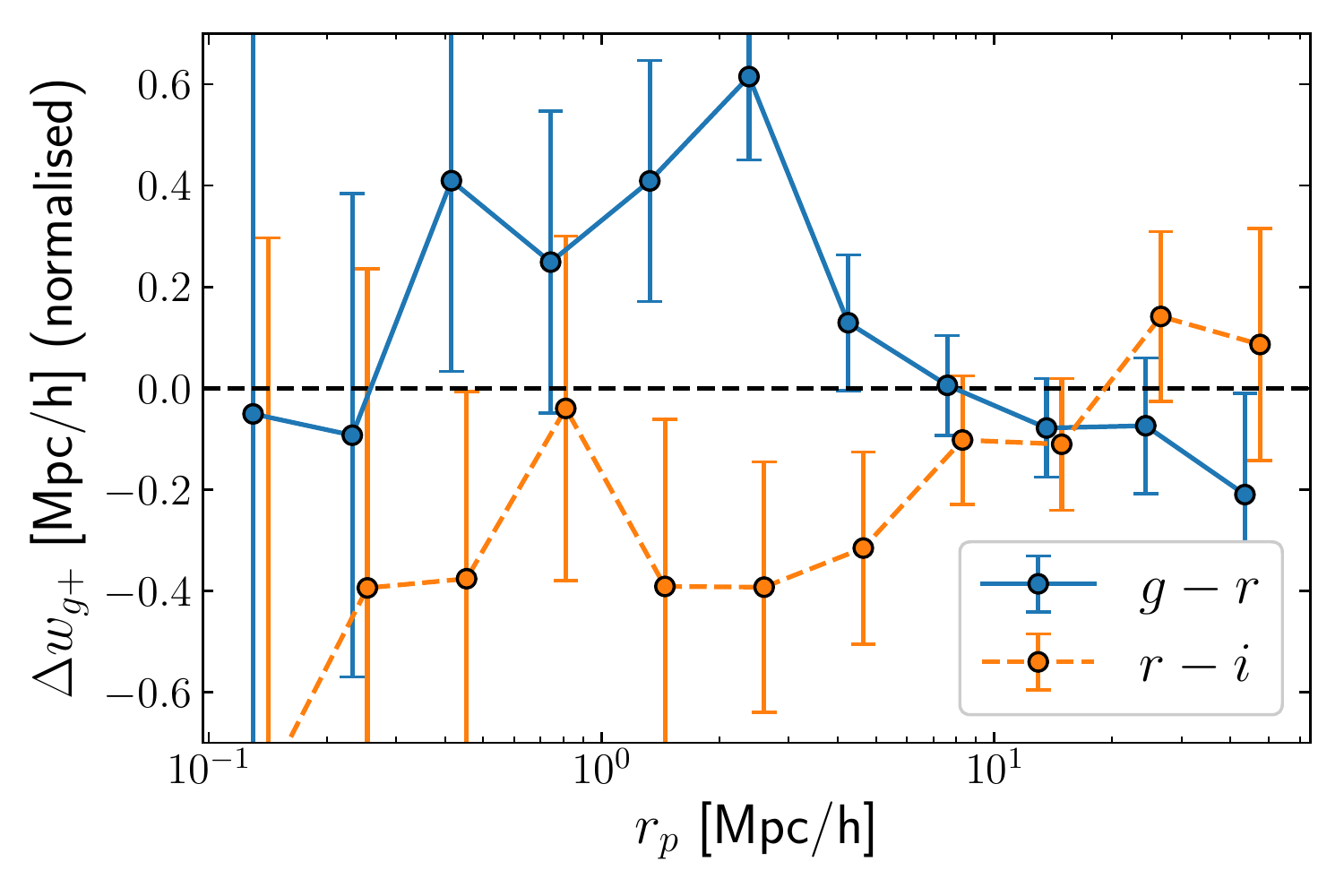}}
	\caption{Difference in projected density-shape correlation function between the $g-r$ and $r-i$ filters (blue and orange points, respectively) but using the normalised ellipticities of galaxies, measured using satellites as shape and density samples. The x-axis shows the transverse separation of galaxy pairs. Points are horizontally displaced for clarity.}
	\label{fig:wgpnorm}
\end{figure}

Another complication can arise from the fact that the PSF is generally larger in the $g$ and $i$-band images than in the $r$-band. Since we are using the same, fixed weight function size on all three bands, we cannot probe exactly the same physical scales of each galaxy in the three bands. However, that would imply that in $g$ and $i$-band images we are probing smaller galaxy scales, where the alignment signal is expected to be smaller, than in the $r$-band images. Contrary to this, we are measuring a higher alignment signal in $g$ and $i$-band. In addition, we have split the galaxy population in galaxy resolution, $R_2>0.9$ and $R_2<0.9$ for high and low resolution galaxies respectively (where $R_2$ was measured in the r-band, see Fig. \ref{fig:R0_histograms}). We see no difference in the IA difference when correlating red satellite shapes and positions, between high and low resolution galaxies. We conclude that the PSF differences in the three broad band image data do not influence our result. 

\section{Conclusions}
\label{sec:Discussion}

In this work we have presented a new shape catalogue for galaxies in the GAMA spectroscopic survey (equatorial fields). Shapes were measured using deep imaging data from the KiDS survey in the three SDSS-like $gri$ broad band filters. This allowed us to compare morphological properties of our galaxy sample between the three filters, as well as investigate whether the galaxy intrinsic alignment signal depends on the wavelength band of the observations. 

The shape measurement method employed is the moment-based \textsc{DEIMOS} method \citep{DEIMOS}. DEIMOS corrects the measured weighted moments using a Taylor expansion of the inverse weight function, essentially using higher order weighted moments to approximate the unweighted moments. In addition, it accounts for the convolution of the galaxy and the PSF, without imposing any prior assumptions on the PSF. In order to calculate the PSF moments at the positions of our galaxies we model the PSF using shapelets \citep{shapeletsI}, with a pipeline already tested on KiDS image data \citep{Kuijken2015}.

Dedicated image simulations were used with two goals in mind: finding the optimal setup for the shape measurement method and characterizing its bias, which can lead to false conclusions about the intrinsic alignment signal, if left uncorrected. We have shown that we can measure ellipticities of galaxies with multiplicative bias of $\sim 0.6\%, 0.4\%$ and $0.8\%$ in the $g$, $r$ and $i$-band images, respectively. The additive bias was found to be consistent with zero.

Ellipticity distributions of galaxies measured in the three $gri$ filters are fairly similar, though the $r$-band shapes appear to peak at a lower ellipticity than the $g$ and $i$-band. The reason for this is partly the lower SNR at which galaxies are observed in $g$ and $i$-band, compared to $r$-band images, but mostly due to galaxies having different morphology across these three bands. Galaxy sizes measured in the $g$-band appear to be larger than sizes measured in the $r$-band. The same is not as evident for sizes in $r$ and $i$-band.

Tests for systematics did not reveal a source for spurious alignments. The physical size of the weight function applied when measuring shapes of galaxies is significantly smaller than the scales to which we measure the positive correlation and cannot account for the measured signal on small scales. Correlation of galaxy position with the cross ellipticity components $w_{g\times}$ is measured to be zero (as expected from symmetry). The PSF model did not seem to induce an artificial signal. 

We have measured the projected correlation function $w_{g+}$ of galaxy positions correlated with the galaxy ellipticities in the three bands and computed its difference among them. We find that the difference $\Delta w_{g+}$ is significantly non-zero, positive between filters $g-r$ and negative for $r-i$. Given that $g$-band is bluer than $r$, and since galaxies generally have colour gradients and build-up hierarchically inside-out, the outer, more blue star population of galaxies will be more prominent in blue filters. Outer regions of a galaxy are also more susceptible to tidal fields and therefore we can expect a stronger alignment signal in blue filters. This can physically explain the positive signal in $g-r$ but the negative difference between the low-z $r-i$ filters is counter-intuitive. From this we conclude that this rather simple interpretation is not enough to explain the observed $\Delta w_{g+}$, and accurate modelling of the galaxy's colour gradients, between $r$ and $i$-band filters, as well as of the alignment signal on these small scales is necessary to understand the observed IA differences. This, however is beyond the scope of this exploratory work. We restrict our analysis to the largest scales ($r_p>6$ Mpc$/h$), which are scales where the intrinsic alignment signal is fit in \citet{Johnston}, we find that the difference in intrinsic alignment signal between bands is consistent with zero. Comparison of the difference to the $w_{g+}$ measured in $r$-band only indicates that our result cannot be neglected; on scales $\sim2$Mpc$/h$ the passband dependence is significant.

We try to find the galaxies responsible for the observed difference in intrinsic alignment signal with wavelength. Splitting the galaxy sample into satellites and centrals we find large values in this difference when we correlate shapes of central or satellite galaxies with positions of satellites. Furthermore, the difference is largest when shapes and positions of red satellites are correlated. The difference is consistent with zero using centrals as the density sample. This fact suggests that the difference in alignment signal does not occur from one galaxy group to another but is a rather short-ranged phenomenon observed among the group galaxies, more accurately traced by using the satellites as density tracers. 

Our analysis suggests that the intrinsic alignment signal depends on the wavelength of observations. Hence, priors of IA on a particular broad band filter should not be used in cosmic shear measurements of a different filter without accounting for this dependence. Also, the IA wavelength dependence seems to be driven by the red satellites in the galaxy sample, and disappears if only central galaxies are considered. Therefore it is important to understand the satellite fraction of a galaxy sample. Intrinsic alignments measured in low redshift samples can be different from samples at high redshift due to a change in the satellite fraction. In addition, galaxy colour gradients change as a function of redshift; in a fixed filter band galaxies at high redshift appear redder than the same galaxies at low redhshift. An apparent redshift dependence of the IA signal can be introduced, solely driven by the change of galaxy colour gradients as a function of redshift.

The dependence of the intrinsic alignment signal with wavelength could provide the ability to study cosmology as well. Similar to what has been proposed in \cite{Elisa}, the alignment signal measured in different broad band filters can be used as a multitracer probe of intrinsic alignments and improve constraints on primordial non-Gaussianity. Building on this, future work will focus in using shape measurements with a radially varying weight function, effectively probing inner and outer regions of galaxies, and quantifying the galaxy scale dependence of the intrinsic alignment signal. Such shape measurements can also be combined to probe the intrinsic alignment contamination to galaxy-galaxy lensing more accurately, as demonstrated in \cite{LeonardMandelbaum}.

\begin{acknowledgements}
We thank Mohammadjavad Vakili and Crist\'obal Sif\'on for useful comments on the text and analysis of this work.
Based on data products from observations made with ESO Telescopes at the La Silla Paranal Observatory under programme IDs 177.A-3016, 177.A-3017 and 177.A-3018, and on data products produced by Target/OmegaCEN, INAF-OACN, INAF-OAPD and the KiDS production team, on behalf of the KiDS consortium. OmegaCEN and the KiDS production team acknowledge support by NOVA and NWO-M grants. Members of INAF-OAPD and INAF-OACN also acknowledge the support from the Department of Physics \& Astronomy of the University of Padova, and of the Department of Physics of Univ. Federico II (Naples). GAMA is a joint European-Australasian project based around a spectroscopic campaign using the Anglo-Australian Telescope. The GAMA input catalogue is based on data taken from the Sloan Digital Sky Survey and the UKIRT Infrared Deep Sky Survey. Complementary imaging of the GAMA regions is being obtained by a number of independent survey programmes including GALEX MIS, VST KiDS, VISTA VIKING, WISE, Herschel-ATLAS, GMRT and ASKAP providing UV to radio coverage. GAMA is funded by the STFC (UK), the ARC (Australia), the AAO, and the participating institutions. The GAMA website is http://www.gama-survey.org/.
HHo and AK acknowledges support from Vici grant 639.043.512, financed by the Netherlands Organisation for Scientific Research (NWO). KK acknowledges support by the Alexander von Humboldt Foundation. MV and HHo acknowledges support from the European Research Council under FP7 grant number 279396 and the Netherlands Organisation for Scientific Research (NWO) through grants 614.001.103. NEC is supported by a Royal Astronomical Society research fellowship. HHi is supported by Emmy Noether (Hi 1495/2-1) and Heisenberg grants (Hi 1495/5-1) of the Deutsche Forschungsgemeinschaft as well as an ERC Consolidator Grant (No. 770935).
\end{acknowledgements}

\bibliographystyle{aa} 
\bibliography{references} 

\begin{appendix}
\section{Tests for systematic errors}
\label{sec:systematics}

In this section we perform tests for systematics that could contaminate the measurement of the correlation functions. We showed that the difference in the alignment signal between broad bands is dominated by satellite galaxies. We have checked that, even though satellites are generally fainter and smaller in size than the general galaxy population, this is not enough to introduce an artificial signal. We see this by splitting the sample into high/low stellar mass galaxies and finding no significant difference in the observed IA signals. Furthermore, the bias does not vary greatly with SNR, with $m\sim3$\% for $\textrm{SNR}\sim60$, which is not enough to explain the observed $\Delta w_{g+}$, especially since only $\sim1$\% of our satellite sample have $\textrm{SNR}_{g,i}<60$. Also, we have checked that the bias in the shape measurement does not depend much on galaxy size either, particularly for our galaxy sample ($m<1.5$ \% in $r$-band simulations for any galaxy half-light radius), and satellite galaxies are, in general, not much smaller in size. In addition, the fact that splitting galaxies into centrals and satellites showcases the ``appearance'' and ``disappearance'' of an IA signal difference, together with the fact that the difference is observed at large galaxy separations (all the way to 2 Mpc$/h$) supports our conclusion that a physical mechanism is the cause of the signal, rather than a systematic error. In the following, we look more closely for such errors.

\subsection{Physical scale of the weight function}

The first thing to consider is the physical scale at which the galaxies are probed with the applied weight function, when measuring galaxy shapes. A large weight function may allow light from nearby galaxies to enter in the measured weighted moments and result in a systematic radial signal being present in the correlation function. The physical size of the weight function's semi-major axis ranges up to 21 kpc$/h$\footnote{Only 0.29\% of our sample has a weight function larger than 21 kpc$/h$.} but with a mean value of 8.8 kpc$/h$. Since this is much smaller than the lowest scales at which the correlation function is measured (100 kpc$/h$) and considering that most of the flux from neighbouring galaxies is masked with segmentation maps, we conclude that the weight function is not large enough to introduce a spurious signal in our data.

\subsection{Galaxy density - cross ellipticity correlation}

The correlation of galaxy positions and the \emph{cross} ellipticity component $\xi_{g\times}$ is expected to be zero due to parity symmetry. Therefore, $w_{g\times}$ serves as a test for systematic errors in our analysis. In Fig. \ref{fig:wgc} we show $w_{g\times}$ measured in the three $gri$ filters. Again, we test if these signals are consistent with zero by performing a $\chi^2$ test against the null-signal hypothesis, taking into account the full shape-noise covariance matrix. It is found that the signals are consistent with zero at a 95\% confidence level with $p$-values equal to 0.26, 0.27 and 0.32 for the $g$, $r$ and $i$-band measurement. Therefore, $w_{g\times}$ revealed no evidence of systematic errors.

\begin{figure}
	\resizebox{\hsize}{!}{\includegraphics{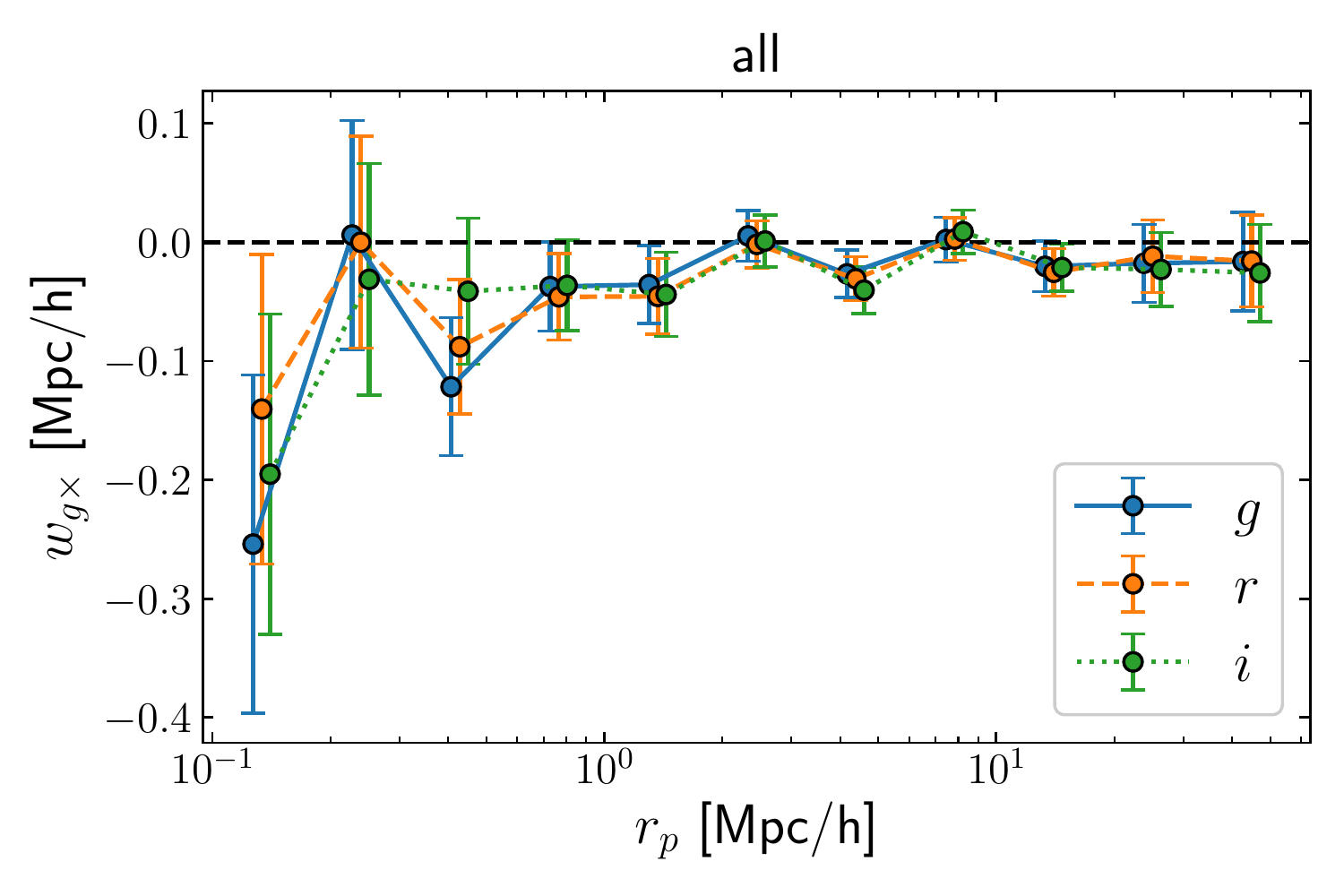}}
	\caption{Projected correlation function between the density and cross ellipticity of all galaxies, for the three filters ($gri$ in blue, orange and green, respectively). The x-axis shows the transverse separation of galaxy pairs. Points are horizontally displaced for clarity.}
	\label{fig:wgc}
\end{figure}

\subsection{PSF shape contamination}

The three-dimensional correlation function $\xi_{g+}$ can be written as a sum of two terms: the first term is the correlation of the galaxy position with the galaxy's intrinsic ellipticity and the second term is proportional to the correlation of the galaxy position with the ellipticity of the PSF, as measured in the position of galaxies \citep[Eq. 24]{Singh}. The latter is expected to be zero since the PSF is not expected to align with galaxy position. We checked that this second term, of the correlation between PSF shapes and galaxy position, is much smaller than the correlation of galaxy shapes and positions (finding a $\sim10^{-3}$ Mpc$/h$ contribution to $w_{g+}$), and does not influence our results.

\section{IA difference for red/blue, high/low redshift galaxies}
\label{sec:redshift_color}
We present here the difference in intrinsic alignment signal between measurements in the $gri$ broad band filters, where we split our galaxy sample into low/high redshift and intrinsically red/blue sub-samples. 

In Fig. \ref{fig:wgpz} we show the difference in alignment between galaxies at low redshift ($z<0.26$) and galaxies at high redshift ($z>0.26$). We notice that for galaxies at high redshift the difference in alignment signal is consistent with zero while galaxies at low redshift exhibit a clear non-zero difference. However, we note that the high redshift sample contains $\sim15,000$ satellites, as opposed to the $\sim33,000$ present in the low redshift sample. We also point out that a measurement in the higher redshift bin is expected to have a lower SNR due to the fact that the volume of this galaxy sample is larger than the volume of the low redshift sample.

In Fig. \ref{fig:wgpcolor} the difference $\Delta w_{g+}$ is shown for intrinsically red and blue galaxies. In both cases the signal is non-zero but the red galaxy population exhibits a stronger difference over larger transverse separations than the blue one. Interestingly intrinsic alignments are expected to be stronger and act on larger scales for red galaxies, compared to blue, according to the linear alignment and tidal torquing models, used commonly to describe intrinsic alignments. 

\begin{figure}
	\resizebox{\hsize}{!}{\includegraphics{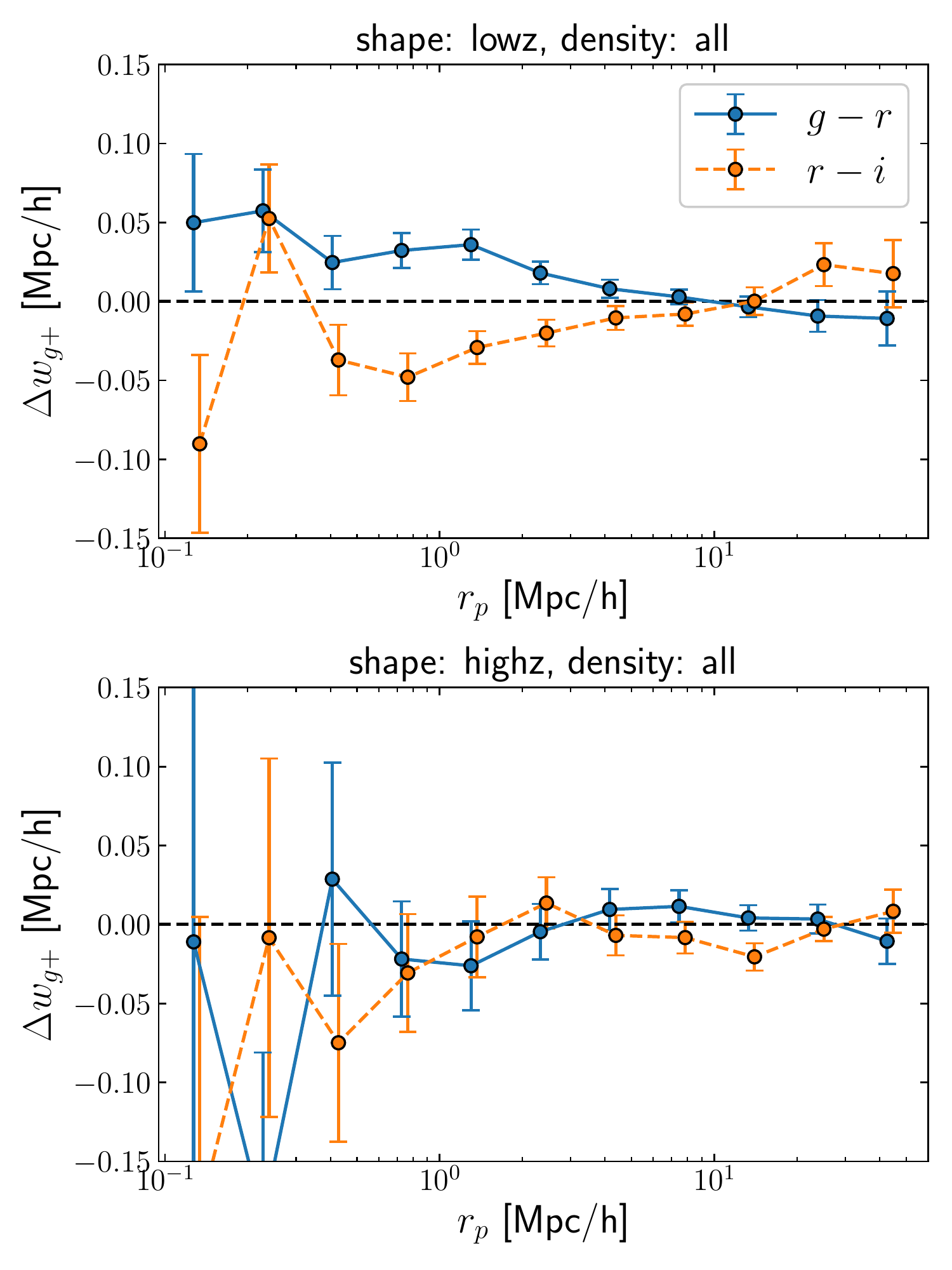}}
	\caption{Difference in projected density-shape correlation function between the $g-r$ and $r-i$ filters (blue and orange points, respectively). The correlations were measured using 
		low (top panel) and high (bottom panel) redshift shape samples ($z<0.26$ and $z>0.26$, respectively) while the density sample contained all galaxies. The x-axis shows the transverse separation of galaxy pairs. Points are horizontally displaced for clarity.}
	\label{fig:wgpz}
\end{figure}

\begin{figure}
	\resizebox{\hsize}{!}{\includegraphics{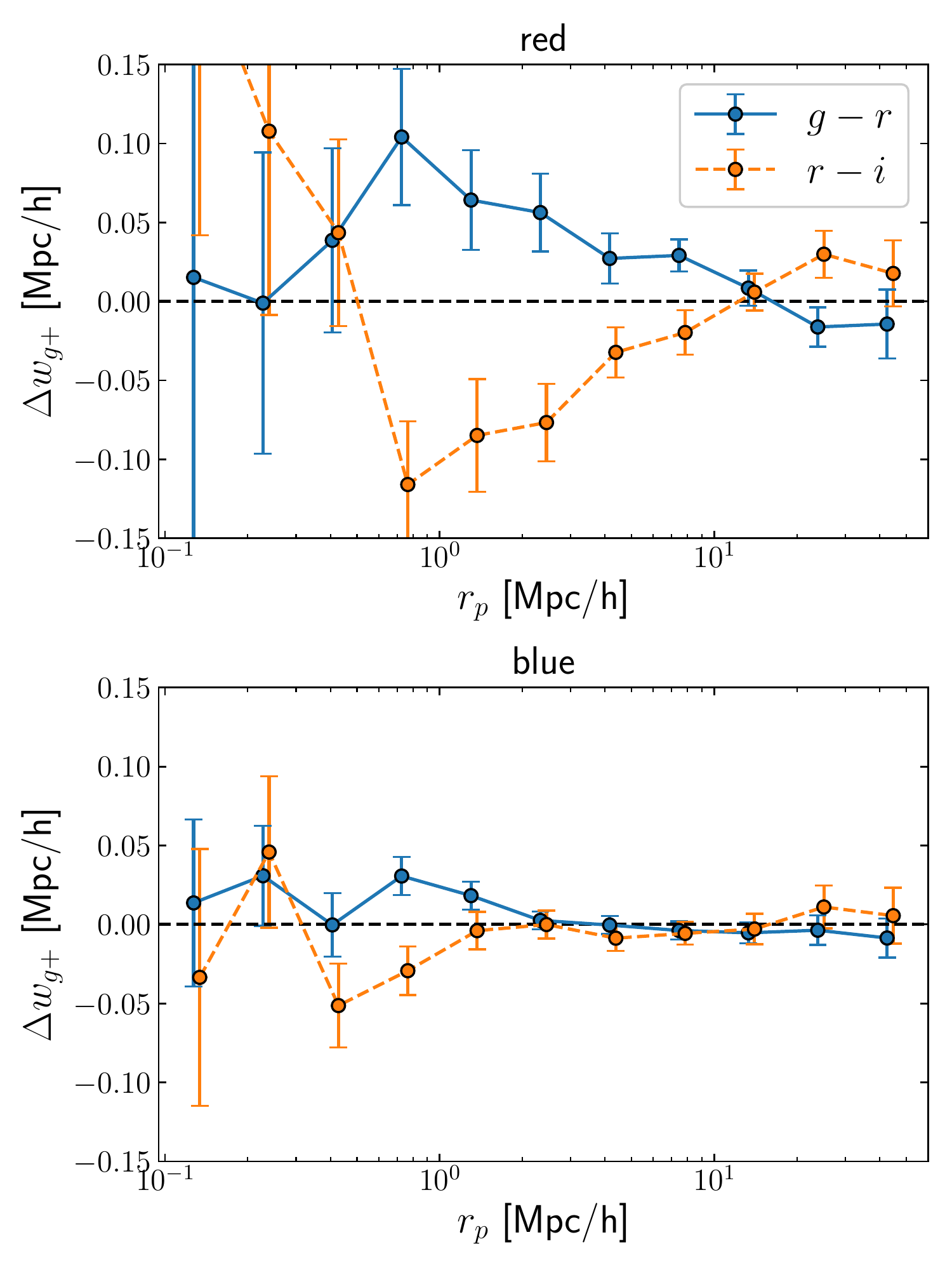}}
	\caption{Difference in projected density-shape correlation function between the $g-r$ and $r-i$ filters (blue and orange points, respectively). The correlations were measured using 
		intrinsically red (top panel) and blue (bottom panel) galaxy sub-samples for both the shape and density field. The x-axis shows the transverse separation of galaxy pairs. Points are horizontally displaced for clarity.}
	\label{fig:wgpcolor}
\end{figure}

\end{appendix}

\end{document}